\documentclass[journal]{IEEEtran}
\usepackage{cite}

\usepackage[dvipsnames]{xcolor}

\ifCLASSINFOpdf
  \usepackage[pdftex]{graphicx}
\else
\fi
\usepackage{amsmath}
\usepackage[caption=false,font=footnotesize]{subfig}
\begin{document}
%


\title{Shaping Frequency Dynamics in Modern Power Systems with Grid-Forming Converters}

%
%

\author{Carlos~Collados-Rodriguez,
        Daniel Westerman Spier,
        Marc~Cheah-Mane,~\IEEEmembership{Member,~IEEE,}
        Eduardo~Prieto-Araujo,~\IEEEmembership{Senior~Member,~IEEE}
        and~Oriol~Gomis-Bellmunt,~\IEEEmembership{Fellow,~IEEE}
        
}
\maketitle

\begin{abstract}
In this paper, frequency dynamics in modern power systems with a high penetration of converter-based generation is analysed. A fundamental analysis of the frequency dynamics is performed to identify the limitations and challenges when the converter penetration is increased. The voltage-source behaviour is found as an essential characteristic of converters to improve the initial frequency derivative of Synchronous Generators (SGs). A detailed small-signal analysis, based on the system's eigenvalues, participation factors and mode shapes, is then performed in a reduced system for different converter penetrations, showing that the flexibility of grid-forming (GFOR) converters as well as the system's inertia reduction may lead to have a more controllable system frequency. First-order frequency responses can be programmed for high converter penetrations, when GFOR operation can impose their dominance over SGs. These results have been validated in the IEEE 118-bus system simulated~in~PSCAD.



\end{abstract}

\begin{IEEEkeywords} 
Grid-forming, frequency dynamics, small-signal analysis.
\end{IEEEkeywords}

%
\IEEEpeerreviewmaketitle

\section{Introduction}
%
%
%
%


\IEEEPARstart{R}{enewable} energy generation is currently increasing in many countries to tackle the effects of climate change. Small islands, like El Hierro (Spain) with a peak load below 10~MW, have already achieved 100\% of instantaneous renewable power during several days \cite{Hoke2021}. Larger islands like Ireland have securely operated the system with a renewable penetration around 75\% and it is planned to increase this value to 95\% by 2030 \cite{Eirgrid2022}. A renewable generation higher than 100\% respect to the local demand has been achieved in Spain, Denmark and Portugal during some hours of the year \cite{Esios,Holttinen2019}, while Italy and Germany have exceeded 80\%. Swedish power system is planned to be fully based in renewable generation by 2040 \cite{InternationalRenewableEnergyAgencyIRENA2020}, while Spain's objective is set for 2050 \cite{GobiernodeEspana.MinisteriodeIndustria2020}. Following these challenging targets, a massive amount of solar and wind power plants are being installed, whose capacity worldwide has been multiplied by eight in just a decade \cite{InternationalRenewableEnergyAgencyIRENA2022}, displacing conventional power plants from the generation mix. Therefore, converter-based resources are replacing Synchronous Generators (SGs) in electrical networks, resulting in a deep transformation of power systems and their operation.

\begin{figure}[b]
	\centering
	\includegraphics[width=1\linewidth]{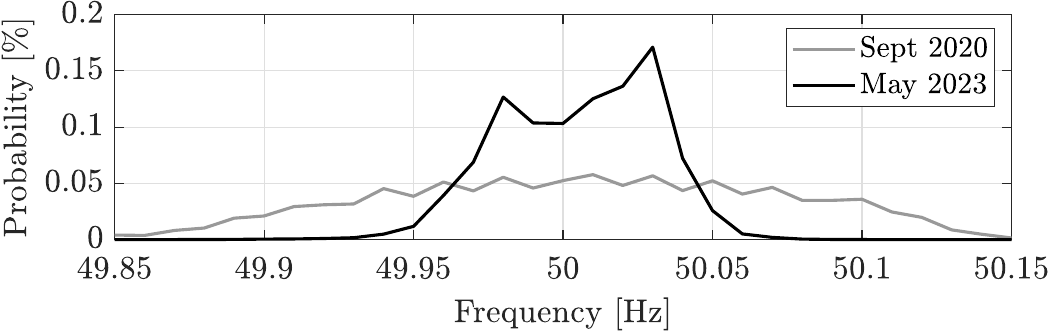}
	\caption{Measured frequency probability in Australia. Data from \cite{AEMO2023b}.} 

	\label{fig:freq_AEMO}
\end{figure}

Voltage-Source Converters (VSCs) are fully controllable devices that might provide higher flexibility to the system, as they do not follow SG's mechanical laws. However, they do not have physical inertia and present a limited short-circuit current, which differs considerably from the conventional SG's behaviour. The inertia reduction due to the SG replacement by power converters poses new challenges for the power system operation. System's inertia is considered nowadays an essential parameter to ensure frequency stability \cite{ENTSO-E2018,ENTSO-E2017}, as it determines the initial Rate of Change of Frequency (RoCoF) after a generation-load imbalance. Several Transmission System Operators (TSOs) estimate the system inertia to guarantee a minimum secure value \cite{ENTSO-E2017, Ashton2015}. In some cases, the installation of synchronous condensers is also being considered to increase the inertia and short-circuit current of the system \cite{Gu2018,ElectraNet2019}. On the other hand, faster active power response of converters can reduce the minimum inertia required to operate the system \cite{Denholm2020}, changing the paradigm of the inertia needs in the system. In October 2020, AEMO required a primary frequency regulation to all converter-based generation \cite{AEMO2021}, helping to improve the frequency deviations in the system. Figure~\ref{fig:freq_AEMO} represents the probability of occurrence of the measured frequency in Australia in September 2020 and May 2023, revealing the reduction in the frequency deviations after including the converter-based generation into the frequency regulation. In addition, an increasing interest in STATCOMs with energy storage is also emerging as an alternative solution to mitigate the impact of load imbalances~\cite{GermanTransmissionSystemOperators2020,Page2022,Meng2022}. 

Although frequency deviations can be mitigated by including the converter-based generation and storage systems in the frequency regulation (as shown in the Australian system in Fig.~\ref{fig:freq_AEMO}), frequency dynamics during large transient events are also a concern, specially when the system's inertia is reduced. In these cases, the frequency dynamics are more dependent on the support provided by converters, which is determined by their control structure. Grid-following (GFOL) VSCs can provide limited support, as their active power injection depends on the frequency measurement, resulting in a delayed response. Additional control strategies have been presented to improve the frequency support of GFOL converters considering also the frequency derivative \cite{Ekanayake2004}. However, the practical implementation of these control schemes can be challenging to apply. Additionally, it has been reported that high GFOL-VSC penetrations can result in system's instability~\cite{NERC2017, CIGRE2016}.

To overcome such stability issues, several grid-forming (GFOR) control structures have been presented \cite{Migrate2017,Rosso2021, Wang2020}. In contrast to the delayed GFOL's response, GFOR VSCs can inject active power instantaneously after a load imbalance, which might help to mitigate the RoCoFs. 
Another important characteristic of GFOR converters is that their frequency no longer depends on physical laws, like the SG's inertia, but just on their control algorithms. This fact opens a new scenario, where GFOR converters can shape the frequency dynamics in power-electronic-dominated power systems. It should also be noted that GFOR converters are self-synchronised units, so it is also necessary to ensure synchronism among all the generators in the system. Therefore, the selected control strategy will play a key role in the future frequency dynamics.

 
Some publications have highlighted that GFOR converters can help to improve frequency dynamics in systems with high penetrations of power electronics. 
The synchronization mechanism between SGs and VSCs is analysed in \cite{Sajadi2022}, where the generation elements are simplified as second-transfer functions. Adjusting the damping component in GFOR converters can improve grid synchronization stability, especially for low-inertia systems. Similar conclusions are derived in \cite{Lasseter2020}, where aggregated models are used to capture the frequency dynamics. A qualitative analysis is provided in \cite{Tayyebi2020}, comparing different synchronization control techniques for GFOR operation. However, a fundamental analysis on how the frequency can evolve in modern power system is not addressed in previous publications. An in-depth analytical and simulation-based study is provided in \cite{Kenyon2023}, where the impact of different droop variants on frequency dynamics is also analysed. It is concluded by means of non-linear simulations that the frequency dynamics progressively tend to a first-order response when the penetration of droop-based GFOR converters is increased.

In this paper, the fundamentals on frequency dynamics are revisited. The impact of the GFOR operation is analysed for different converter penetrations and converter control parameters. The study is based on a small-signal analysis, providing details about the oscillation modes (frequency and damping), participation factors and mode shapes. This small-signal analysis allows to fully characterise the frequency dynamics, which are later validated via EMT simulations implemented in PSCAD. In particular, the following contributions are considered:
\begin{itemize}
    \item A comprehensive study about frequency dynamics in modern low-inertia power systems is provided, revealing that the VSC's dominance can be an opportunity to improve frequency dynamics. It is demonstrated that high penetrations of GFOR converters allow to shape the frequency dynamics, which can be designed to follow the desired response, \textit{e.g.} a first-order response.


    \item The dominant modes of frequency dynamics are identified by small-signal analysis. These dominant modes have been classified as \textit{Synchronisation mode} and \textit{Global mode}, which fully describe the frequency behaviour in a power system with multiple generation units.
    
    \item A sensitivity analysis of the GFOR VSC's synchronisation control parameters is performed in a reduced system, identifying their impact on the frequency dynamics for different converter penetrations. 
    \item The results obtained in the reduced system are validated with a full non-linear model of the IEEE 118-bus system, implemented in PSCAD, confirming that it is possible to reshape the frequency dynamics also in large power systems with GFOR converters.

    \item The voltage-source behaviour is shown as an essential characteristic from converters to improve frequency dynamics with high penetrations of power electronics.
    
\end{itemize}

\section{Fundamentals on frequency dynamics}
\subsection{Frequency dynamics in SG-based power systems}
Conventional power systems have their foundations in synchronous generators. The SGs performance is based on a rotating magnetic field, typically produced by the current injected by the exciter into the rotor windings, which induces an electro-magnetic force in the stator windings proportional to its angular frequency. The stator windings are then connected to the rest of the system. Therefore, SGs perform as voltage sources behind an impedance \cite{Kundur1994}.

\begin{figure}[t]
	\centering
	
	\subfloat[]{\includegraphics[width=\linewidth]{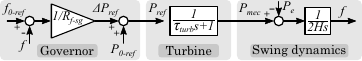}
	\label{fig:SG_freq_scheme}}

    \subfloat[]{\includegraphics[width=\linewidth]{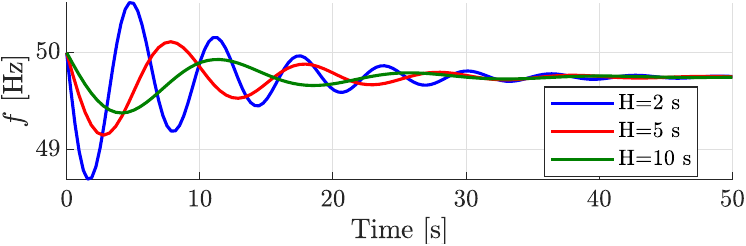}
    \label{fig:freq_inertia_effect}}
    
	\caption{Frequency dynamics of a conventional power system: (a) Reduced-order model; (b) Effect of SG inertia.}
	\label{fig:schemes}
\end{figure}

This voltage-source behavior of the SGs allows them to naturally react to load imbalances, increasing or reducing their active and reactive power based on the system's needs. Consequently, the SGs' electrical torque changes as a result of active power demand/generation variations. Due to the slower dynamics of the turbine when applying the mechanical torque, the extra electrical power injected to the network can only be extracted from the kinetic energy stored in the machines, resulting in a frequency deviation after a load imbalance. The frequency dynamics in conventional power systems can be described by the SGs' swing equation: 
\begin{equation}\label{eq:SG_freq}
    \dfrac{\text{d}f}{\text{d}t}=\dfrac{1}{2H}(P_{mec}-P_e)
\end{equation}
where $f$ is the SG's frequency, $P_{mec}$ is the mechanical power provided by the prime mover, $P_e$ is the electrical power injected by the electrical machine and $H$ is the SG's inertia. The dynamics of the prime mover can be approximately described as the following first-order function \cite{Kundur1994}:
\begin{equation} \label{eq:turb}
    \dfrac{P_{mec}(s)}{P_{ref}(s)}=\dfrac{1}{\tau_{turb}s+1}
\end{equation}
where $\tau_{turb}$ is the time constant of the turbine and $P_{ref}$ is the reference power generated by the governor, which is usually implemented as a droop control with a characteristic $R_{f-sg}$ 
(measured in Hz/MW or pu/pu): 
\begin{equation}\label{eq:SG_droop}
    P_{ref}=P_{0-ref}+\frac{1}{R_{f-sg}}(f_{0-ref}-f)
\end{equation}
where $P_{0-ref}$ is the active power setpoint at the frequency reference $f_{0-ref}$ (typically the nominal frequency). Combining \eqref{eq:SG_freq}, \eqref{eq:turb} and \eqref{eq:SG_droop} leads to the model shown in Fig.~\ref{fig:SG_freq_scheme}. Hence, frequency dynamics after a load change in conventional power systems 
can be characterised by the following second-order transfer function:
\begin{equation} \label{eq:SG_freq_2}
    \dfrac{f(s)}{P_{e}(s)}=\dfrac{\tau_{turb}s+1}{2H\tau_{turb}s^2+2Hs+1/R_{f-sg}}
\end{equation}

The turbine's time constant $\tau_{turb}$ is in the order of several seconds \cite{Kundur1994}, which is considerably slower compared to the electrical dynamics. Therefore, during the first seconds after a load change, while the mechanical torque has not still reached the new setpoint, only the inertia of the SGs can limit the RoCoF. This can be observed in \eqref{eq:SG_freq},  where the frequency derivative is inversely proportional to the inertia constant $H$. Reducing the system inertia leads to higher values of RoCoF, as shown in Fig.~\ref{fig:freq_inertia_effect}, which can increase the probability of SG trips or cause hard damages to SG-based power plants~\cite{ENTSO-E2016}. 

Since the inertia plays a key role on keeping the RoCoF within proper limits during the first seconds, it is known as \textit{inertial response} \cite{Tielens2020}. It should be noted that inertial response refers to the source of energy, which is provided by the kinetic energy of the machines, while the operation principle of the electrical power response is given by the SGs' voltage-source behavior.

\subsection{Frequency dynamics in presence of grid-following VSCs}
\begin{figure}[t]
	\centering
	
	\subfloat[]{\includegraphics[width=\linewidth]{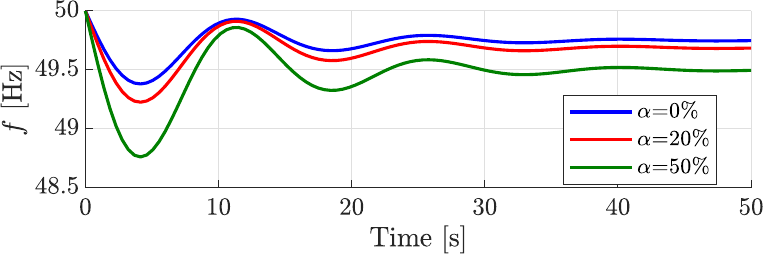}
	\label{fig:freq_alpha_gfol_no_sup}}

    \subfloat[]{\includegraphics[width=\linewidth]{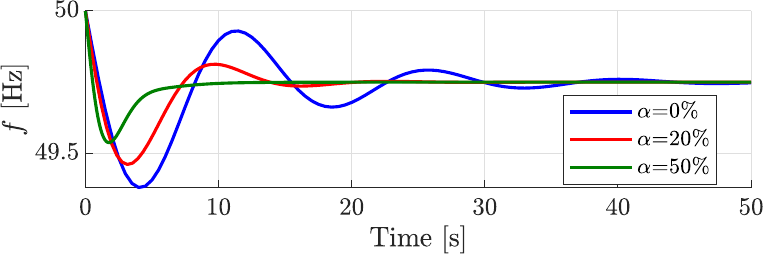}
    \label{fig:freq_alpha_gfol_droop}}
	\caption{Frequency dynamics with an $\alpha$ penetration of GFOL VSC: (a) without frequency support ($\beta=0$); (b) with proportional frequency support ($\beta=1$).}
	\label{fig:freq_gfol}
\end{figure}

Replacing SGs by VSC-based generation can impact the frequency dynamics since converters' operating principles differ from the electrical machines. GFOL VSCs operate as current sources, whose references can be calculated from outer control layers to support the grid, \textit{i.e.} frequency and voltage support~\cite{Rosso2021}. As a result, GFOL VSCs do not respond intrinsically to load imbalances, leaving this function to SGs. Therefore, for the same load imbalance, SGs must inject a higher power when their capacity is replaced by converter-based generation. To include the effects caused by the addition of GFOL VSCs into the power system, the swing equation presented in \eqref{eq:SG_freq} can be modified as: 
\begin{equation}\label{eq:SG_freq_gfol}
    \dfrac{\text{d}f}{\text{d}t}=\dfrac{1}{2H(1-\alpha)}((1-\alpha)P_{mec}-P_d+\alpha \beta P_{gfol})
\end{equation}
where $H$ is the original inertia of the system with 100\% SGs, $P_d$ is the active power demand, $P_{gfol}$ is the active power injected by the converters, $\alpha$ is the converter penetration ($\alpha\in$ [0,1)) and $\beta$ is the proportion of VSCs that provides frequency support ($\beta\in$ [0,1]). Increasing $\alpha$ would reduce the total system inertia, challenging the frequency stability. If converters do not participate in the frequency regulation, the system frequency would experience larger deviations for any VSC penetration, as shown in Fig.~\ref{fig:freq_alpha_gfol_no_sup}. In addition, the steady-state frequency is also affected as less generation units participate in the primary regulation. Therefore, it is clear that converters should participate in the frequency regulation.

To provide frequency support to the system, GFOL converters modify their active power reference according to the measured system's frequency. A proportional droop control is typically used for this purpose \cite{Morren2006a}, resulting in the following response: 
\begin{equation}\label{eq:freq_support_gfol}
    \dfrac{P_{gfol}(s)}{f(s)}=\dfrac{1/R_{f-gfol}}{(\tau_{pll}+\tau_f+\tau_P)s+1}=\dfrac{1/R_{f-gfol}}{\tau_{p-gfol}s+1}
\end{equation}
where $R_{f-gfol}$ is the $P$-$f$ droop characteristic, and $\tau_{pll}$, $\tau_f$ and $\tau_{p-gfor}$ are the time constants of the Phase-Locked Loop (PLL), frequency filter (if exists) and active power control (if exists) respectively. These three time constants can be combined to obtain the overall time constant of the active power injected by the converter after a frequency disturbance, $\tau_{p-gfol}$. Combining \eqref{eq:turb}, \eqref{eq:SG_freq_gfol} and \eqref{eq:freq_support_gfol}, a reduced-order model for the frequency dynamics which considers frequency support from GFOL converters can be obtained: 
\begin{equation} \label{eq:freq_gfol}
    \dfrac{f(s)}{P_{d}(s)}=\dfrac{\left(\tau_{turb}s+1\right)\left(\tau_{p-gfol}s+1\right)}{a_3s^3+a_2s^2+a_1s+a_0}
\end{equation} 
where 
\begin{align}
    a_3 = & 2H(1-\alpha)\tau_{turb}\tau_{p-gfol}\\
    a_2 = & 2H(1-\alpha)(\tau_{turb}+\tau_{p-gfol})\\
    a_1 = & (2H+1/R_{f-sg})(1-\alpha)+\tau_{turb}\alpha\beta/ R_{f-gfol} \\
    a_0 = & (1-\alpha)/R_{f-sg}+\alpha/ R_{f-gfol}
\end{align} 

\begin{figure}[t]
	\centering
	
	\subfloat[]{\includegraphics[width=0.49\linewidth]{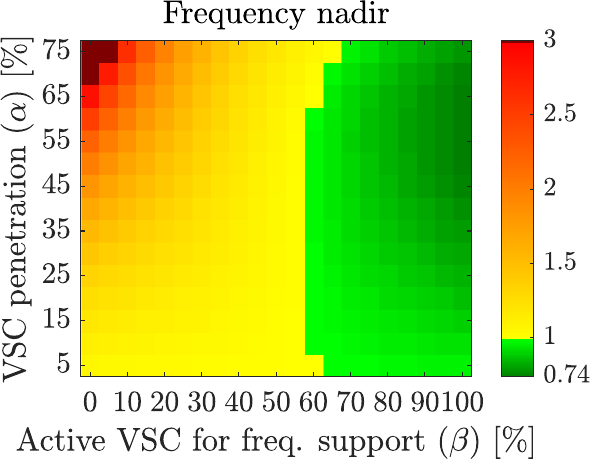}
	\label{fig:nadir_alpha_beta_gfol}}
    \subfloat[]{\includegraphics[width=0.49\linewidth]{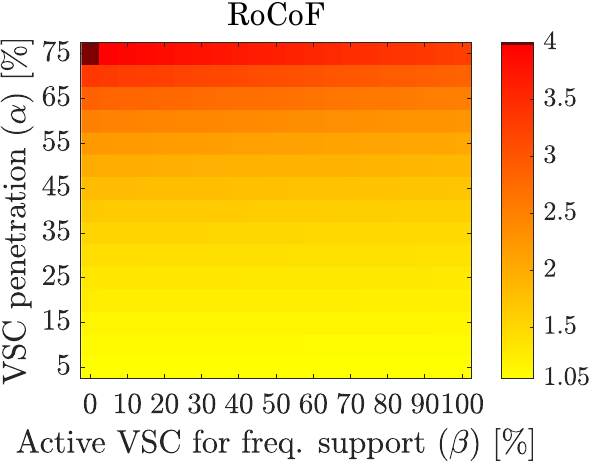}
    \label{fig:rocof_alpha_beta_gfol}}

	\caption{Frequency nadir and RoCoF evolution for different values of VSC penetration ($\alpha$) and proportion of active VSCs in frequency support~($\beta$).}
	\label{fig:freq_alpha_beta}
\end{figure}

Converters can regulate their active power injection in hundreds of milliseconds (in this case, $\tau_{p-gfol}$ has been set to 250~ms), while conventional SG governors and turbines require a few seconds to provide mechanical torque to the electrical machine. Therefore, converters can help to enhance frequency dynamics as they have more flexibility in controlling their active power compared to SG governors, as shown in Fig.~\ref{fig:freq_alpha_gfol_droop}. 
The minimum value of the frequency during the transient, also known as frequency nadir, can be improved by the fast converter response. However, the reduction of the system inertia caused by the introduction of converter-based generation results in a higher initial RoCoF due to the non-intrinsic response of GFOL converters. GFOL VSCs operate as current sources that react after frequency deviations to compensate them. This response is significantly different from the voltage-source behavior shown by SGs, which provide power instantaneously after a load change. Fig.~\ref{fig:freq_alpha_beta} shows the frequency nadir and the RoCoF obtained for different values of VSC penetration ($\alpha$) and proportion active VSC for frequency support ($\beta$) respect to the 100\% SG case. It can be observed that for $\beta$ higher than 70\%, the nadir can be improved for any VSC penetration. Nevertheless, the RoCoF is always larger than the base case. 
Therefore, to limit the initial RoCoF, converters must provide instantaneous response after active power imbalances. In other words, they must behave as voltage sources. 

\subsection{Frequency dynamics in presence of grid-forming VSCs}
Grid-forming operation is characterised by controlling both the magnitude and phase of the voltage either at the converter's terminals or at the Point Of Connection (POC) \cite{Rosso2021}. Consequently, GFOR VSCs operate as voltage sources, providing all the power demanded by the rest of the system while their capacity is not exceeded and the energy resource is available. Unlike SGs, a stand-alone GFOR converter can operate at constant frequency even after load variations. The control loop that ensures this operation is the voltage control (or an equivalent control structure), which forces the converter to inject the required current to regulate the voltage at the desired level. In this sense, GFOR VSCs show a superior performance compared to SGs, whose frequency dynamics depend on the mechanical torque provided by the turbine. 

When connecting a GFOR converter to other generation units, such as SGs or other GFOR VSCs, a specific control loop is necessary to ensure synchronization. Several implementations have been presented for this purpose in the literature \cite{Rosso2021}, in which the two main used are the Virtual Synchronous Machine (VSM) and droop-based control. For the sake of simplicity, in this manuscript the droop-based approach has been selected as the option for the synchronization loop.  

\begin{figure}[t]
	\centering
	\includegraphics[width=0.7\linewidth]{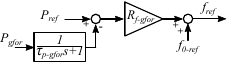}
	\caption{Power-frequency droop control in GFOR operation.}
	\label{fig:gfor_droop_scheme}
\end{figure}
\begin{figure}[t]
	\centering
	\includegraphics[width=1\linewidth]{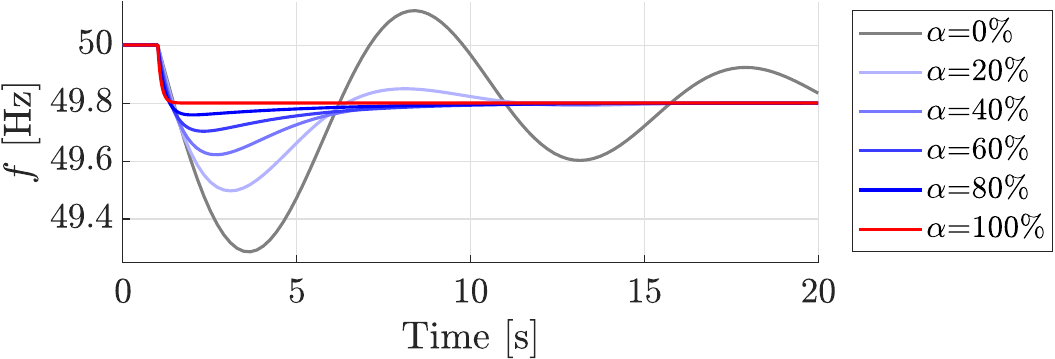}
	\caption{Frequency dynamics for different $\alpha$ of droop-based GFOR VSC.} 
	\label{fig:GFOR_global_mode}
\end{figure}

Considering a stand-alone converter now provided with a power-frequency droop control, as the one shown in Fig.~\ref{fig:gfor_droop_scheme}, the characteristic transfer function of the frequency is:
\begin{equation} 
    \dfrac{f(s)}{P_{gfor}(s)}=\dfrac{R_{f-gfor}}{\tau_{p-gfor}s+1}
\end{equation} 
where $R_{f-gfor}$ is the $P$-$f$ characteristic and $\tau_{p-gfor}$ is the time-constant of the measured active power filter. In contrast to SGs, which exhibit a second-order response, the frequency dynamics of GFOR converters employing the $P$-$f$ droop control approach follow a first-order response after a load change. Therefore, considering a power system with both SGs and droop-based GFOR VSCs, both first-order and second-order responses are expected for the frequency dynamics. Similarly to \eqref{eq:freq_gfol} for GFOL, the equation which describes the frequency response in presence of droop-based GFOR converters is:
\begin{equation} \label{eq:freq_gfor}
    \dfrac{f(s)}{P_{d}(s)}=\dfrac{\tau_{turb}s+1}{b_2s^2+b_1s+b_0}
\end{equation}
where 
\begin{align}
    b_2 = & \left[2H(1-\alpha)+\tau_{p-gfor}\alpha/R_{f-gfor}\right]\tau_{turb}\\
    b_1 = & 2H(1-\alpha)+(\tau_{turb}+\tau_{p-gfor})\alpha/R_{f-gfor}\\
    b_0 = & (1-\alpha)/R_{f-sg}+\alpha/R_{f-gfor}
\end{align}
Fig.~\ref{fig:GFOR_global_mode} shows the frequency response in \eqref{eq:freq_gfor} for different values of converter penetration ($\alpha$), while $\tau_{p-gfor}$ has been set to 100~ms. It can be observed that increasing $\alpha$ provides higher damping to the system frequency, reaching first order dynamics for high values of $\alpha$. Therefore, the dominant dynamics may depend on the specific proportion of SGs and VSCs in the generation mix as well as the control values set in the converter control. This will be discussed in the next sections using a full detailed model, including all the dynamics that are not considered in this preliminary analysis.

\section{Frequency dynamics in modern power systems} \label{sec:analysis}
\begin{figure}[t]
	\centering
	\subfloat[]{\includegraphics[width=0.8\linewidth]{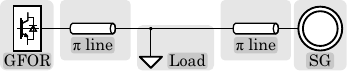}
	\label{fig:SG_GFOR_scheme}}
    
	\subfloat[]{\includegraphics[width=\linewidth]{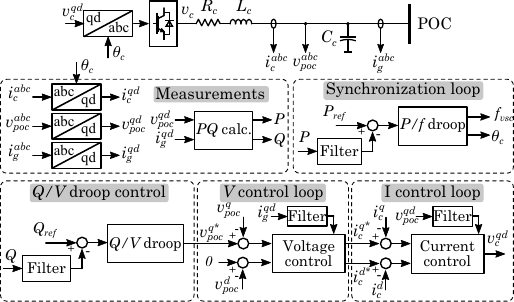}
	\label{fig:GFOR_control_scheme}}
	\caption{System under study: (a) Complete system; (b) Grid-forming control.}
	\label{fig:system_scheme}
\end{figure}
This section investigates the frequency dynamics in modern power systems with high penetration of renewables, where the converter control can play a key role on the system performance. In such futuristic scenario, the VSCs are assumed to operate in GFOR control mode. To such end, a reduced system has been considered, shown in Fig.~\ref{fig:SG_GFOR_scheme}, which is composed by a SG, a GFOR VSC and a load, capturing all the essential dynamics of the main components. The GFOR control, represented in Fig.~\ref{fig:GFOR_control_scheme}, is based on cascaded control loops for the AC current and AC voltage, both implemented in the $qd$ reference frame. The synchronisation loop is based on a $P$-$f$ droop, while a $Q$-$V$ droop is implemented to obtain the AC voltage reference at POC. Two different cases have been evaluated to analyse the effect of the converter's penetration in the system frequency dynamics:
\begin{itemize}
    \item Case 1: Low VSC penetration (20\%) 
    \item Case 2: High VSC penetration (80\%)
\end{itemize}
The main system parameters are shown in Table~\ref{tab:system_parameters}. Additionally, a sensitivity analysis of the VSC's synchronisation control parameters has been performed to illustrate their impact on the frequency dynamics. All the cases are studied through a small-signal analysis and EMT simulations. In particular, system poles, Participation Factors (PFs) and mode shapes are studied for the small-signal analysis, while a load change increase of 10\%, \textit{i.e.} 40~MW, is considered as the disturbance for the non-linear simulations. For the small-signal analysis, the computation of a state-space model of the complete system is required, which can be built following the methodology presented in \cite{Collados-Rodriguez2020}.

Firstly, the PFs have been examined to identify the oscillation modes that participate in the frequency dynamics, \textit{i.e.} poles associated to the SG frequency $\omega_{SG}$ and the VSC frequency $\omega_{VSC}$. The values of the PFs have been normalised between 0 (no participation of a variable in a pole) and 1 (maximum participation). A greyscale has been used to represent the PF matrices, in which white colour refers to no participation and black colour indicates maximum participation. By analysing the characteristics of the participating poles, the dominant dynamics on the frequency response can be identified. Mode shapes are the right eigenvectors of the state-space matrix and show how an oscillation mode is mapped in the different state variables. They have been used in power system analysis to identify coherent groups of SGs in electro-mechanical oscillations \cite{Klein1991}. Finally, the EMT simulation results of the non-linear models are shown to validate previous small-signal analysis.

\begin{table}[t]
\caption{Main system parameters}
\label{tab:system_parameters}
\begin{tabular}{lcccc}
\hline
\textbf{Parameter}         & \textbf{Symbol} & \textbf{Unit} & \textbf{Case 1} & \textbf{Case 2} \\ \hline
VSC penetration            & $\alpha$ & \%        & 20            & 80            \\
Nominal frequency   & $f_n$   & Hz       & 50           & 50          \\
Rated SG power             & $S_{sg}$  & MVA      & 400         & 100         \\
SG droop characteristic    & $R_{f-sg}$ & \%       & 5            & 5             \\
SG droop gain              & $k_{f-sg}$  & MW/Hz    & 160      & 40        \\
Rated VSC power            & $S_{vsc}$  & MVA     & 100         & 400         \\
VSC droop characteristic    & $R_{f-gfor}$   & \%     & 5            & 5            \\
VSC droop gain              & $k_{f-gfor}$  & MW/Hz     & 40   & 160        \\
SG active power ref.  & $P^*_{sg}$ & MW      & 320           & 80          \\
VSC active power ref. & $P^*_{vsc}$ & MW    & 80           & 320          \\
Initial power demand       & $P_{load}$  & MW    & 400          & 400         \\ \hline
\end{tabular}
\end{table}

\subsection{Case 1: Low VSC penetration (20\%)}

\begin{table*}[htpb]
\caption{Case 1: oscillation modes which participate in frequency dynamics and their characteristics}
\label{tab:case1_oscillation_modes}
\centering
\begin{tabular}{cccc|cccc|cccc|cccc}
\hline
\multicolumn{4}{c|}{\textbf{$\tau_{p-gfor}$ = 10 ms}}                                                  & \multicolumn{4}{c|}{\textbf{$\tau_{p-gfor}$ = 100 ms}} & 
\multicolumn{4}{c|}{\textbf{$\tau_{p-gfor}$ = 1 s}}   &
\multicolumn{4}{c}{\textbf{$\tau_{p-gfor}$ = 5 s}}  \\ \hline
\textbf{Pole}        & \textbf{$f_n$ {[}Hz{]}} & \textbf{$\xi$} & \textbf{Var.}         & \textbf{Pole} & \textbf{$f_n$ {[}Hz{]}} & \textbf{$\xi$} & \textbf{Var.}   &                   
\textbf{Pole}        & \textbf{$f_n$ {[}Hz{]}} & \textbf{$\xi$} & \textbf{Var.}         & \textbf{Pole} & \textbf{$f_n$ {[}Hz{]}} & \textbf{$\xi$} & \textbf{Var.} \\ \hline
15-16                & 11.38           & 0.44                           & $\omega_{VSC}$        & 16-17         & 4.58                  & 0.08                           & $\omega_{VSC}$ & 19-20                & 1.76                 & 0.04                           & \begin{tabular}[c]{@{}c@{}}$\omega_{SG}$,\\ $\omega_{VSC}$\end{tabular}       & 21-22         & 1.20                  & 0.04                           & \begin{tabular}[c]{@{}c@{}}$\omega_{SG}$,\\ $\omega_{VSC}$\end{tabular}\\

29-30                & 0.085           & 0.78                           & $\omega_{SG}$         & 29-30         & 0.08           & 0.77                           & $\omega_{SG}$ & 30-31                & 0.069       & 0.71                           & \begin{tabular}[c]{@{}c@{}}$\omega_{SG}$,\\ $\omega_{VSC}$\end{tabular}         & 30-31         & 0.047                 & 0.63                           & \begin{tabular}[c]{@{}c@{}}$\omega_{SG}$,\\ $\omega_{VSC}$\end{tabular}\\ \hline
\end{tabular}

\end{table*}
\begin{figure*}[t]
	\centering
	\subfloat[$\tau_{p-gfor}$ = 10 ms]{\includegraphics[width=0.265\linewidth]{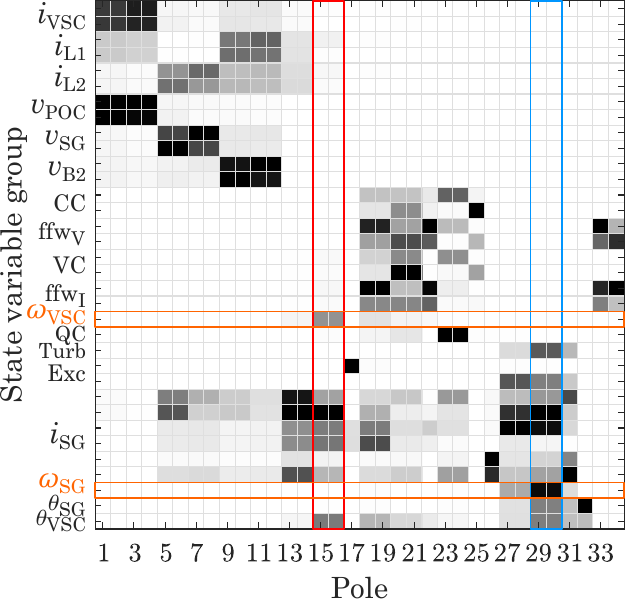}
	\label{fig:PF_400_tau_10ms}}
    \subfloat[$\tau_{p-gfor}$ = 100 ms]{\includegraphics[width=0.225\linewidth]{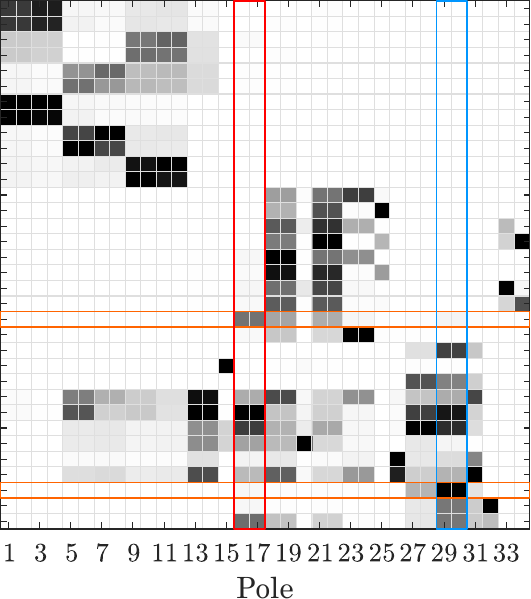}
    \label{fig:PF_400_tau_100ms}}
	\subfloat[$\tau_{p-gfor}$ = 1 s]{\includegraphics[width=0.225\linewidth]{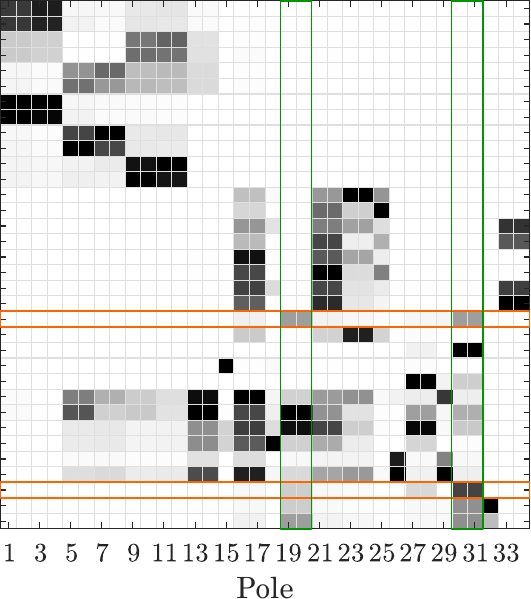}
	\label{fig:PF_400_tau_1s}}
    \subfloat[$\tau_{p-gfor}$ = 5 s]{\includegraphics[width=0.252\linewidth]{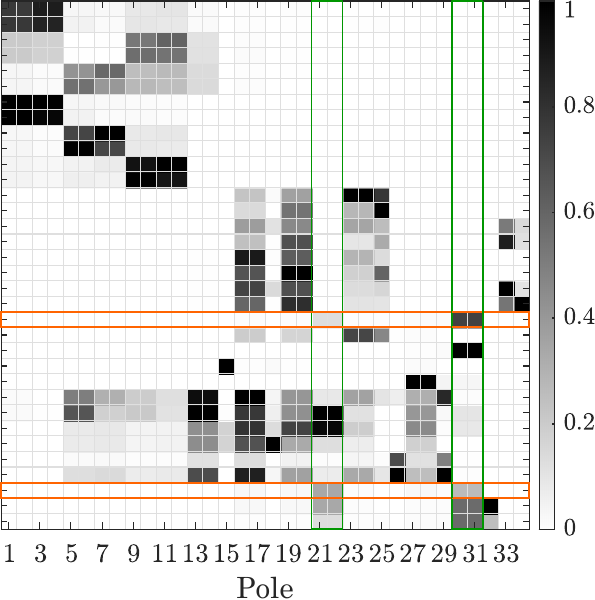}
    \label{fig:PF_400_tau_5s}}

	\caption{Case 1: PF matrices for different values of $\tau_{p-gfor}$. Color code that indicates participation: red = VSC; blue = SG; green = VSC + SG.}
	\label{fig:case1_PF}
\end{figure*}

Case 1 considers a converter penetration of 20\%, consisting of a system with a 400-MVA SG and a 100-MVA GFOR VSC. The system power demand is 400~MW, which is proportionally shared by the SG (320~MW) and the VSC (80~MW). The impact of the active power droop filter of the converter has been examined for four different values of $\tau_{p-gfor}$ (10 ms, 100 ms, 1 s and 5 s), while the droop characteristic ($R_{VSC}$) is kept constant and equal to 5\% (40 MW/Hz).

\begin{figure}[t]
	\centering
	\includegraphics[width=1\linewidth]{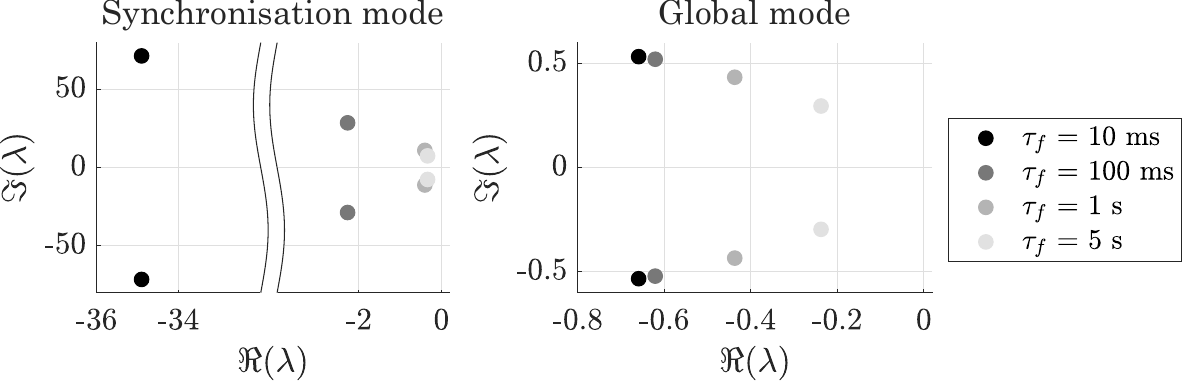}

	\caption{Case 1: Dominant poles of frequency dynamics.}
	\label{fig:case1_poles}

\end{figure}

\begin{figure}[t]
	\centering
	\includegraphics[width=1\linewidth]{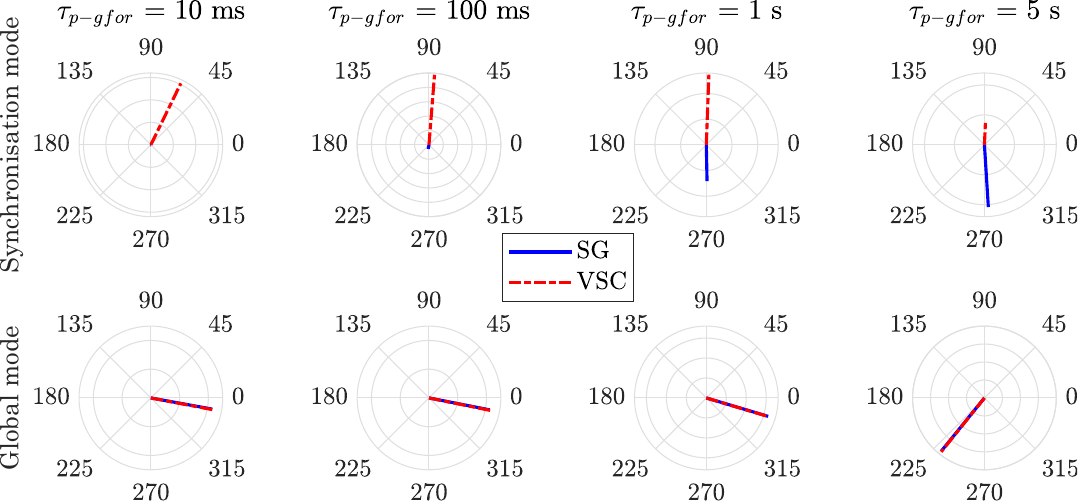}
	\caption{Case 1: Mode shapes of dominant poles.}
	\label{fig:case1_mode_shapes}
\end{figure}

The PF matrices are shown in Fig.~\ref{fig:case1_PF}. It can be observed that the converter frequency $\omega_{VSC}$ and the SG frequency $\omega_{SG}$ participate in different poles for low values of $\tau_{p-gfor}$ (10 ms and 100 ms), as shown by the dark coloured squares in Figs.~\ref{fig:PF_400_tau_10ms} and~\ref{fig:PF_400_tau_100ms}. The poles which participate in the $\omega_{VSC}$ dynamics are highlighted in red, while those corresponding to $\omega_{SG}$ are represented in blue. Therefore, the dynamics of $\omega_{VSC}$ and $\omega_{SG}$ are decoupled for these values of $\tau_{p-gfor}$ (10 ms and 100 ms). The characteristics of the participating poles, natural frequency ($f_n$) and damping ratio ($\xi$), are shown in Table~\ref{tab:case1_oscillation_modes}. Modes associated to $\omega_{VSC}$ are faster than those related to the SG. 
The dynamics of $\omega_{VSC}$ are dominated by the pole pair 15-16 for $\tau_{p-gfor}$ = 10~ms and the pole pair 16-17 for $\tau_{p-gfor}$ = 100~ms. Therefore, it is expected that $\omega_{VSC}$ from the VSC synchronisation loop oscillates at 11.38~Hz and 4.58~Hz for each case, respectively. PF matrices indicate that $\omega_{SG}$ is not affected by these oscillations, as it does not participate in these modes (depicted in white colour for poles 15-16 in Figs.~\ref{fig:PF_400_tau_10ms} and poles 16-17 in Fig~\ref{fig:PF_400_tau_100ms}). 
The participating poles of $\omega_{SG}$ are 29-30, which have an oscillation frequency of 0.08~Hz and a damping ratio around 0.77. This oscillation mode is very similar in both cases, as the VSC is much faster than these dynamics and does not affect the SG frequency response. For slower time constants of the VSC's active power filter ($\tau_{p-gfor}$ = 1~s and 5~s), the dynamics of $\omega_{VSC}$ are in a similar time range compared to $\omega_{SG}$, leading to oscillation modes associated to both variables. These modes are depicted in green in Figs.~\ref{fig:PF_400_tau_1s} and~\ref{fig:PF_400_tau_5s}. Therefore, the two dominant pole pairs are now related to both $\omega_{SG}$ and $\omega_{VSC}$. 

These two dominant modes can be classified as \textit{Synchronisation mode} and \textit{Global mode}. The Synchronisation mode refers to the fast dynamics found during the synchronisation between the SG and the GFOR VSC. The Global mode can be understood as the frequency dynamics of center of inertia, which is slower compared to the Synchronisation mode. Fig.~\ref{fig:case1_poles} represents the poles related to the Synchronisation and Global modes. It can be observed that both dynamics follow a second-order response which is highly affected by the value of $\tau_{p-gfor}$.

The mode shapes of the Synchronisation and Global modes are represented in Fig.~\ref{fig:case1_mode_shapes} for all the values of $\tau_{p-gfor}$. It is observed that the Synchronisation mode mainly affects the VSC for $\tau_{p-gfor}$ = 10~ms and 100~ms. The converter is much faster than the SG, whose frequency dynamics are more rigid compared to the converter's ones. Therefore, the VSC makes all the efforts to adapt its frequency to remain synchronised with the dominating SG, oscillating at 11.38~Hz and 4.58~Hz for $\tau_{p-gfor}$ = 10~ms and 100~ms respectively. For higher values, both SG and VSC present counter-phase mode shapes, indicating that both units swing against each other during the synchronisation period with an oscillation frequency of 1.76~Hz and 1.2~Hz for $\tau_{p-gfor}$ = 1~s and 5~s respectively. Therefore, the synchronising task is shared between both units. The mode shapes of the Global mode show that both units are in phase and follow the same dynamics for the values of $\tau_{p-gfor}$ considered. 

\begin{table*}[t]
\centering
    \caption{Case 1: Main characteristics of the frequency response}
    \label{tab:case1_freq_chars}
\begin{tabular}{lcccc}\hline
                                    & \textbf{$\tau_{p-gfor}$ = 10 ms} & \textbf{$\tau_{p-gfor}$ = 100 ms} & \textbf{$\tau_{p-gfor}$ = 1 s} & \textbf{$\tau_{p-gfor}$ = 5 s} \\ \hline
\textbf{SG freq. nadir}             &  49.65 Hz                        & 49.65 Hz                          &  49.67 Hz                      & 49.73 Hz                        \\
\textbf{SG freq. RoCoF}    &  -0.36 Hz/s                      & -0.35 Hz/s                        &  -0.28 Hz/s                    & -0.12 Hz/s                       \\
\textbf{Average VSC power (500 ms)} & 4.39 MW                          & 5.92 MW                           & 11.47 MW                       & 25.38 MW                        \\
\textbf{Average VSC power (1 s)}    & 6.86 MW                          & 7.66 MW                           & 12.90 MW                       & 24.84 MW                        \\
\textbf{Average VSC power (2 s)}    & 10.10 MW                         & 10.54 MW                          & 14.44 MW                       & 23.93 MW                        \\
\textbf{Average VSC power (10 s)}   & 9.88 MW                          & 9.94 MW                           & 10.53 MW                       & 13.51 MW \\ \hline                      
\end{tabular}
\end{table*}

\begin{figure*}[t]
	\centering
	
	\subfloat[$\tau_{p-gfor}$ = 10 ms]{\includegraphics[width=0.24\linewidth]{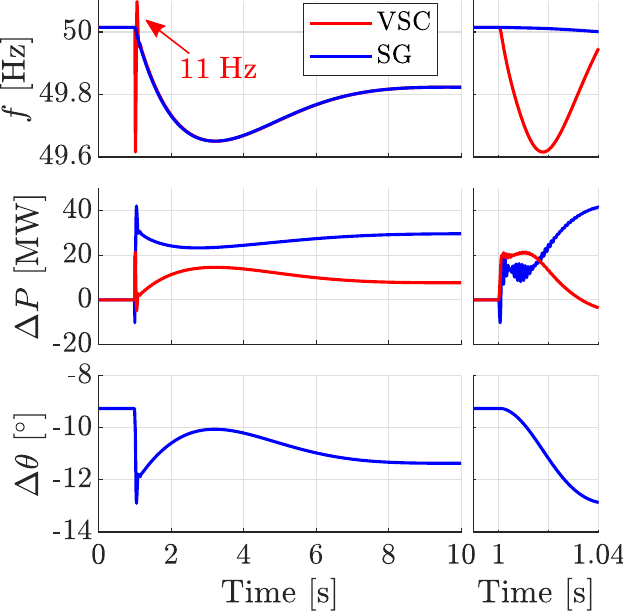}
	\label{fig:sim_resutls_400_tau_10ms}}
    \subfloat[$\tau_{p-gfor}$ = 100 ms]{\includegraphics[width=0.24\linewidth]{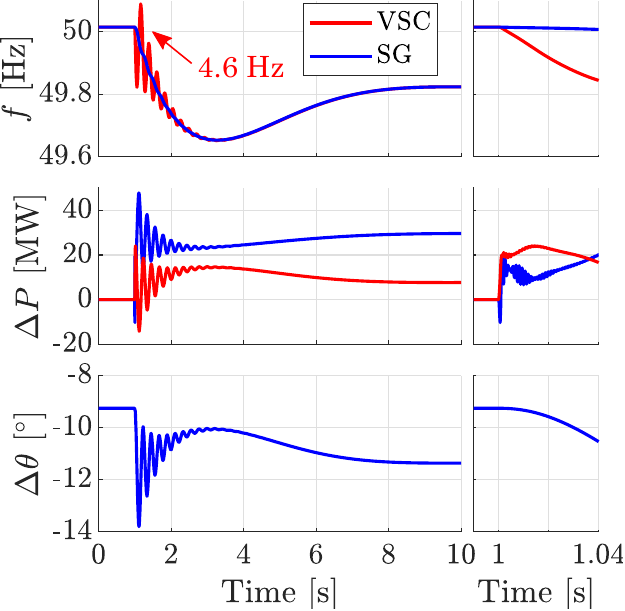}
    \label{fig:sim_resutls_400_tau_100ms}}
    \subfloat[$\tau_{p-gfor}$ = 1 s]{\includegraphics[width=0.24\linewidth]{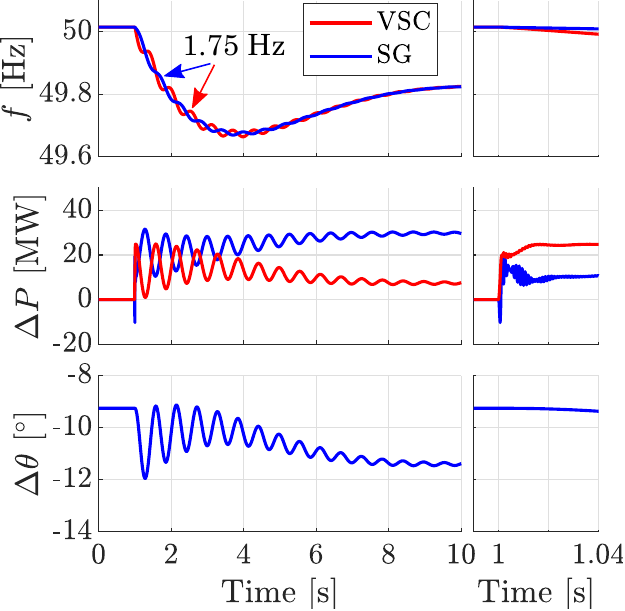}
    \label{fig:sim_results_400_tau_1s}}
    \subfloat[$\tau_{p-gfor}$ = 5 s]{\includegraphics[width=0.24\linewidth]{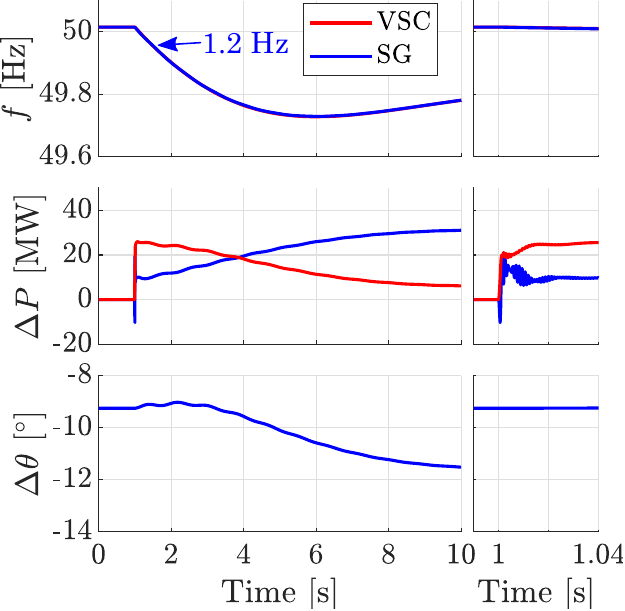}
    \label{fig:sim_results_400_tau_5s}}
    
	\caption{Case 1: Non-linear simulation results.}
	\label{fig:case1_sim_results}
\end{figure*}

The previous small-signal analysis has been validated through EMT simulations of the non-linear model. The results of VSC and SG frequencies, active power increase ($\Delta P$) and angle difference ($\Delta \theta$) are depicted in Fig.~\ref{fig:case1_sim_results} for all values of $\tau_{p-gfor}$ (10 ms, 100 ms, 1 s and 5 s), which confirm the conclusions drawn from the small-signal analysis. It can also be observed the voltage-source behavior of the GFOR VSC, which responds instantaneously to the load change. The VSC active power increase is very similar for the four values of $\tau_{p-gfor}$ during the first 10~ms approximately, confirming that the responsible of voltage-source behavior is the voltage control. After this time, the VSC frequency deviates from the steady-state value according to the time constant $\tau_{p-gfor}$ implemented, initiating the synchronisation and power sharing process between the converter and the generator. Table~\ref{tab:case1_freq_chars} shows the main characteristics of the frequency response: SG frequency nadir, maximum SG RoCoF (measured using a moving average with a time window of 500~ms) and the average VSC active power for different time windows (500~ms, 1~s, 2~s and 10~s) after the disturbance. It can be observed that the RoCoF is within the operational limits of 0.5-1~Hz/s defined by several TSOs \cite{ENTSO-E2016, EirGrid2019, Nationalgrid2019}. In addition, higher values of $\tau_{p-gfor}$ lead to better results of frequency nadir and especially the RoCoF. However, for such nadir and RoCoF improvements, a large active power injection is required from the converter, which shares most of the load increase even when its rated power is much lower, potentially exceeding its capabilities or available resource. 

Finally, some key observations can be drawn for Case 1:
\begin{itemize}
    \item The GFOR VSC acts as a voltage source, responding to load changes in a few milliseconds.
    \item The VSC active power variation is the disturbance for the synchronisation loop, which modifies the internal frequency of the VSC according to the time constant of the active power filter.
    \item The information provided by the small-signal analysis, through system poles, PFs and mode shapes, can be used to fully characterise the frequency response.
    \item Low values of $\tau_{p-gfor}$ (10~ms and 100~ms) lead to very low contribution of the converter to the SG's inertial response, but ensures fast synchronisation. The RoCoF mainly depends on the SG inertia. 
    \item High values of $\tau_{p-gfor}$ (1~s) results in a higher VSC's contribution to the SG's inertial response, which helps to reduce the initial RoCoF. Oscillations against other self-synchronised voltage sources, like SGs, are expected. Since the two element are very stiff, they oscillate until they reach synchronism.
    \item For very high values of $\tau_{p-gfor}$ (5~s), the converter provides most of the power required during the transient, imposing its frequency dynamics, closer to a first-order response. Nevertheless, the converter capacity can be exceeded.
\end{itemize}

\subsection{Case 2: High VSC penetration (80\%)}
\begin{figure*}[t]
	\centering
	\subfloat[$\tau_{p-gfor}$ = 10 ms]{\includegraphics[width=0.265\linewidth]{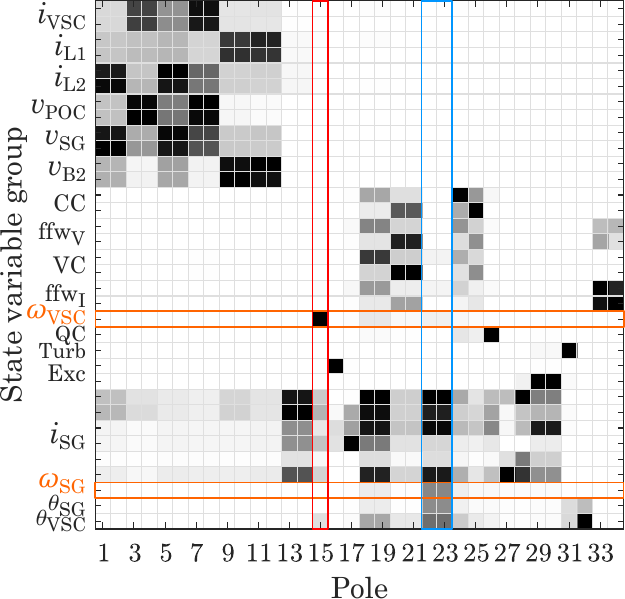}
	\label{fig:PF_100_tau_10ms}}
    \subfloat[$\tau_{p-gfor}$ = 100 ms]{\includegraphics[width=0.225\linewidth]{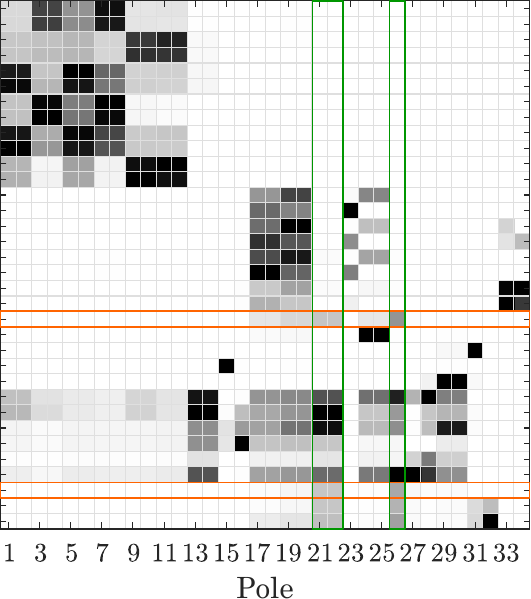}
    \label{fig:PF_100_tau_100ms}}
	\subfloat[$\tau_{p-gfor}$ = 1 s]{\includegraphics[width=0.225\linewidth]{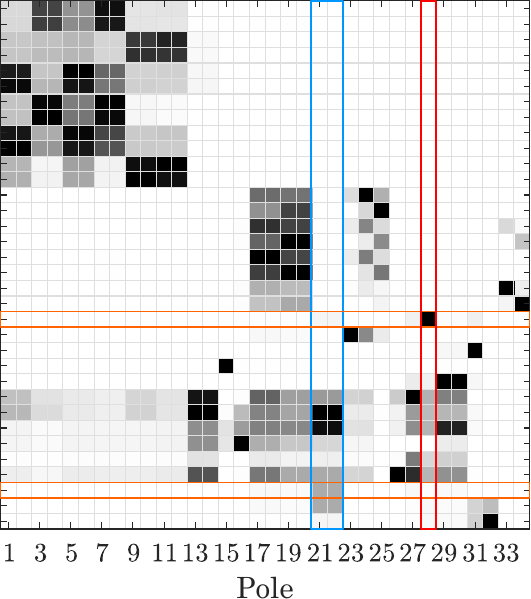}
	\label{fig:PF_100_tau_1s}}
    \subfloat[$\tau_{p-gfor}$ = 5 s]{\includegraphics[width=0.252\linewidth]{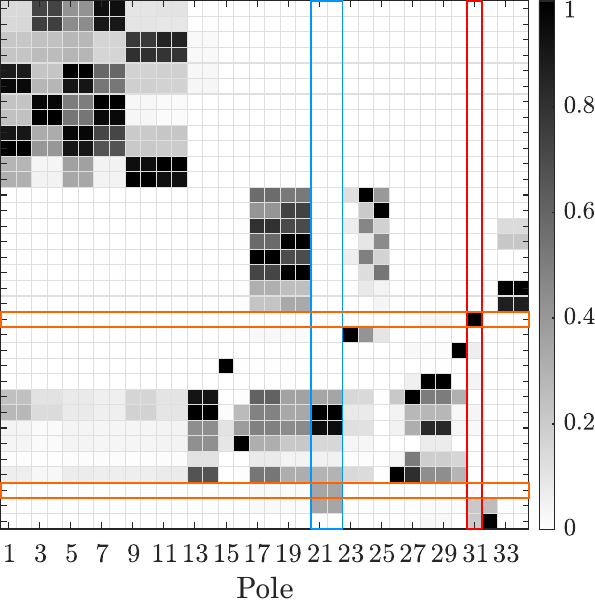}
    \label{fig:PF_100_tau_5s}}

	\caption{Case 2: PF matrices for different values of $\tau_{p-gfor}$. Color code that indicates participation: red = VSC; blue = SG; green = VSC + SG.}
	\label{fig:case2_PF}
\end{figure*}

\begin{figure}[t]
	\centering
	\includegraphics[width=1\linewidth]{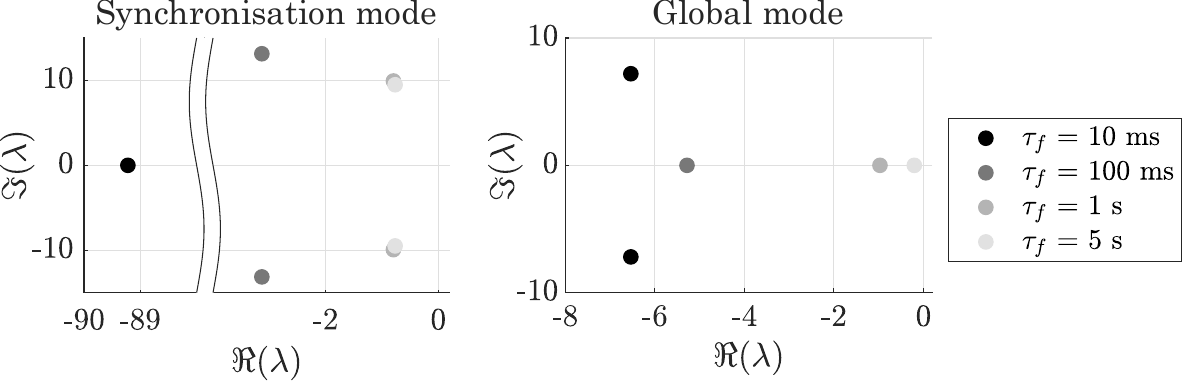}
	\caption{Case 2: Dominant poles of frequency dynamics.}
	\label{fig:case2_poles}
\end{figure}

The same analysis is performed for Case 2, where the converter penetration is increased to 80\%. Therefore, the SG rated power is 100~MVA while the GFOR VSC's size is 400~MVA. PF matrices, pole diagrams of the Synchronisation and Global modes and their characteristics are shown in Fig.~\ref{fig:case2_PF}, Fig.~\ref{fig:case2_poles} and Table~\ref{tab:case2_oscillation_modes}, respectively. It can be observed that the Global mode is in the order of ten times faster compared to Case~1. Additionally, it should be noticed the presence of first-order responses in the frequency dynamics due to the larger size of the VSC, which can impose its dynamics to the Global frequency response. This behaviour differs from Case~1, where both Synchronisation and Global modes followed second-order functions. 

\begin{table*}[htpb]
\caption{Case 2: oscillation modes which participate in frequency dynamics and their characteristics}
\label{tab:case2_oscillation_modes}
\centering
\begin{tabular}{cccc|cccc|cccc|cccc}
\hline
\multicolumn{4}{c|}{\textbf{$\tau_{p-gfor}$ = 10 ms}}                                                  & \multicolumn{4}{c|}{\textbf{$\tau_{p-gfor}$ = 100 ms}} & 
\multicolumn{4}{c|}{\textbf{$\tau_{p-gfor}$ = 1 s}}   &
\multicolumn{4}{c}{\textbf{$\tau_{p-gfor}$ = 5 s}}  \\ \hline
\textbf{Pole}        & \textbf{$f_n$ {[}Hz{]}} & \textbf{$\xi$} & \textbf{Var.}         & \textbf{Pole} & \textbf{$f_n$ {[}Hz{]}} & \textbf{$\xi$} & \textbf{Var.}   &                   
\textbf{Pole}        & \textbf{$f_n$ {[}Hz{]}} & \textbf{$\xi$} & \textbf{Var.}         & \textbf{Pole} & \textbf{$f_n$ {[}Hz{]}} & \textbf{$\xi$} & \textbf{Var.} \\ \hline
15                   & 89.22                & 1                               & $\omega_{VSC}$        & 21-22         & 2.09                  & 0.23                           & \begin{tabular}[c]{@{}c@{}}$\omega_{SG}$,\\ $\omega_{VSC}$\end{tabular} & 21-22                & 1.58                 & 0.08                           & $\omega_{SG}$      & 21-22         & 1.51                  & 0.08                           & $\omega_{SG}$\\
22-23                & 1.14                  & 0.67                            & $\omega_{SG}$         & 26         & 5.28                & 1                              & \begin{tabular}[c]{@{}c@{}}$\omega_{SG}$,\\ $\omega_{VSC}$\end{tabular} & 28                & 0.974                & 1                           & $\omega_{VSC}$         & 31         & 0.22                 & 1                           & $\omega_{VSC}$\\ \hline
\end{tabular}
\end{table*}

\begin{figure}[t]
	\centering
	\includegraphics[width=1\linewidth]{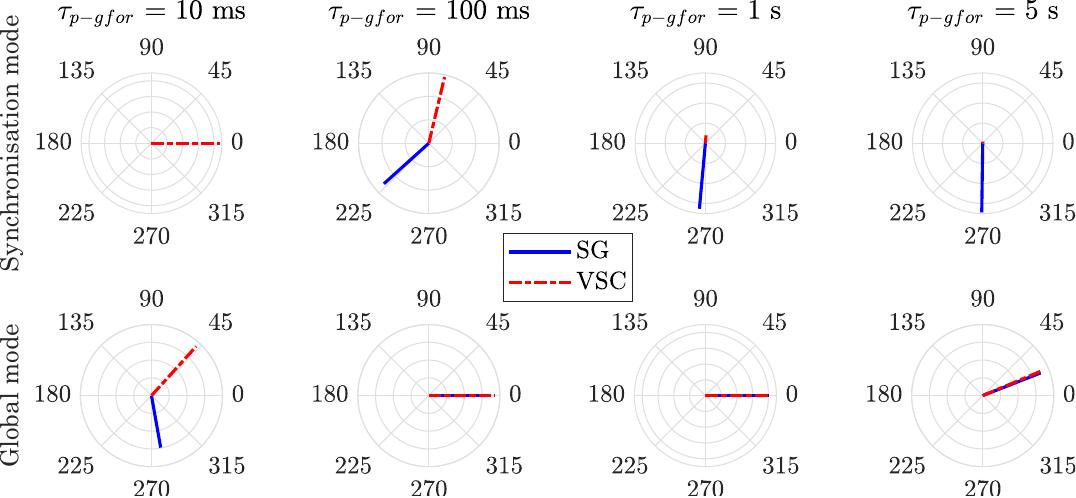}
	\caption{Case 2: Mode shapes of dominant poles.}
	\label{fig:case2_mode_shapes}
\end{figure}

Analysing the particular responses of the VSC and SG frequencies for different values of $\tau_{p-gfor}$, it can be observed that the Global mode is still ruled by a second-order function when  $\tau_{p-gfor}$ = 10~ms (see the complex pole pair of the Global mode in Fig.~\ref{fig:case2_poles}), corresponding to poles 22-23 in Fig.~\ref{fig:PF_100_tau_10ms} and Table~\ref{tab:case2_oscillation_modes}, with $f_n$ = 1.14~Hz and $\xi$ = 0.67. Unlike Case~1, the Synchronisation mode now follows a first-order response with a time constant of 0.011~s (1$/f_n$ = 1$/$89.22~Hz = 0.011~s), which agrees with $\tau_{p-gfor}$. This can also be observed in the real pole of the Synchronisation mode in Fig.~\ref{fig:case2_poles}. Although the converter's rating is higher than the SG's one, the VSC is still much faster than the SG. Thus, the VSC adapts its frequency during the synchronisation period while the Global mode is still influenced by the SG inertia. When $\tau_{p-gfor}$ is increased to 100~ms, the frequency response is fully modified. PFs matrices in Fig.~\ref{fig:PF_100_tau_100ms} show that both $\omega_{VSC}$ and $\omega_{SG}$ participate in the Synchronisation and Global modes, which now follow second and first-order responses respectively, as represented by the complex pole pair and real pole, respectively, in Fig.~\ref{fig:case2_poles}. During the synchronisation, the VSC and the SG swing against each other, as it is depicted by mode shapes of the Synchronisation mode in Fig.~\ref{fig:case2_mode_shapes}. The inertia reduction makes the SG more sensitive to suffer from electro-mechanical oscillations, as the generator is more flexible. Regarding the response of the Global mode, its time constant is 0.19~s (1$/$5.28~Hz), that is slightly higher than $\tau_{p-gfor}$, revealing that the SG inertia still plays a role in the Global mode dynamics. Larger values of $\tau_{p-gfor}$ (1~s and 5~s) make the VSC's frequency more rigid, forcing the SG to adapt its frequency to reach synchronisation. This is observed in poles 21-22 in Table~\ref{tab:case2_oscillation_modes}, which show that the SG suffers from very low damped oscillations of 1.5-1.6~Hz. This is the opposite behaviour of Case~1 with low value of $\tau_{p-gfor}$, where the converter took all the synchronisation efforts due to the slower dynamics of the SG. The dominance of the VSC is also visible in the Global mode, which follows a first-order dynamics of 1.03~s (1$/$0.974~Hz) for $\tau_{p-gfor}$ = 1 s and 4.54~s (1$/$0.22~Hz) for $\tau_{p-gfor}$ = 5~s. This is a key difference between Case~2 and Case~1, which shows that droop-based GFOR VSCs can enable first-order frequency dynamics in modern power systems with high penetration of converters.

\begin{table*}[]
\centering
    \caption{Case 2: Main characteristics of the frequency response}
    \label{tab:case_2_freq_chars}
\begin{tabular}{lcccc}\hline
                                    & \textbf{$\tau_{p-gfor}$ = 10 ms} & \textbf{$\tau_{p-gfor}$ = 100 ms} & \textbf{$\tau_{p-gfor}$ = 1 s} & \textbf{$\tau_{p-gfor}$ = 5 s} \\ \hline
\textbf{SG freq. nadir}             &   49.77 Hz           &    49.79 Hz     &  49.79 Hz                              &  49.80 Hz                                \\
\textbf{SG freq. RoCoF}    &    -0.44 Hz/s                              &    -0.44 Hz/s                               &         -0.15 Hz/s                        &  -0.08 Hz/s                               \\
\textbf{Average VSC power (500 ms)} & 24.94 MW                          & 26.20 MW                           & 31.41 MW                       & 34.23 MW                        \\
\textbf{Average VSC power (1 s)}    & 28.91 MW                          & 29.16 MW                           & 31.67 MW                       & 33.81 MW                        \\
\textbf{Average VSC power (2 s)}    & 30.91 MW                         & 31.06 MW                          & 32.44 MW                       & 34.47 MW                        \\
\textbf{Average VSC power (10 s)}   & 32.59 MW                          & 32.64 MW                           & 33.20 MW                       & 35.12 MW \\ \hline                      
\end{tabular}
\end{table*}
\begin{figure*}[htpb]
	\centering
	
	\subfloat[$\tau_{p-gfor}$ = 10 ms]{\includegraphics[width=0.24\linewidth]{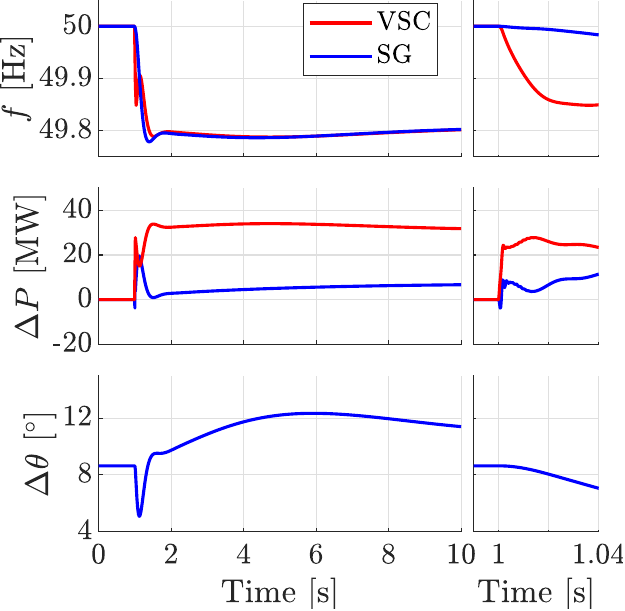}
	\label{fig:sim_resutls_100_tau_10ms}}
    \subfloat[$\tau_{p-gfor}$ = 100 ms]{\includegraphics[width=0.24\linewidth]{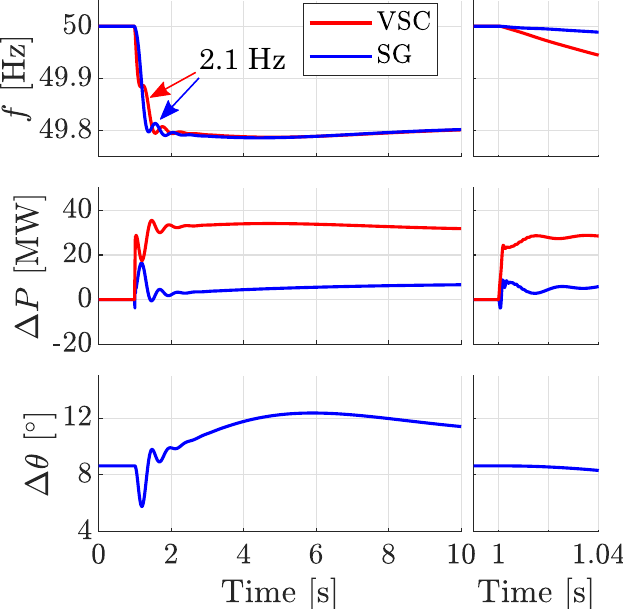}
    \label{fig:sim_resutls_100_tau_100ms}}
    \subfloat[$\tau_{p-gfor}$ = 1 s]{\includegraphics[width=0.24\linewidth]{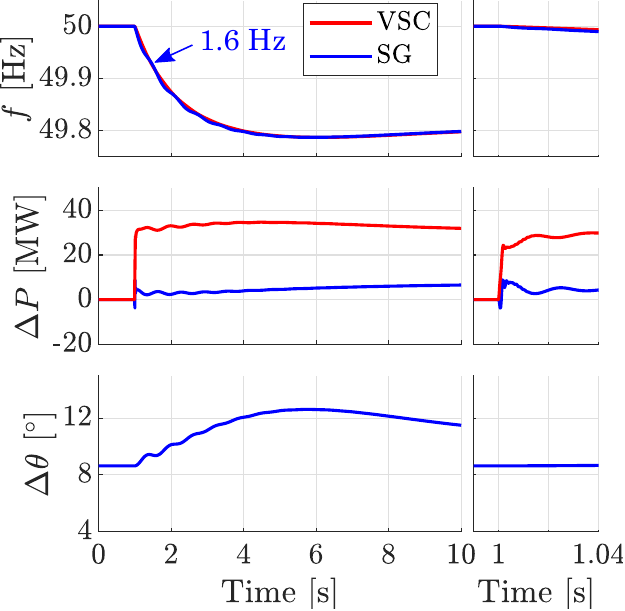}
    \label{fig:sim_results_100_tau_1s}}
    \subfloat[$\tau_{p-gfor}$ = 5 s]{\includegraphics[width=0.24\linewidth]{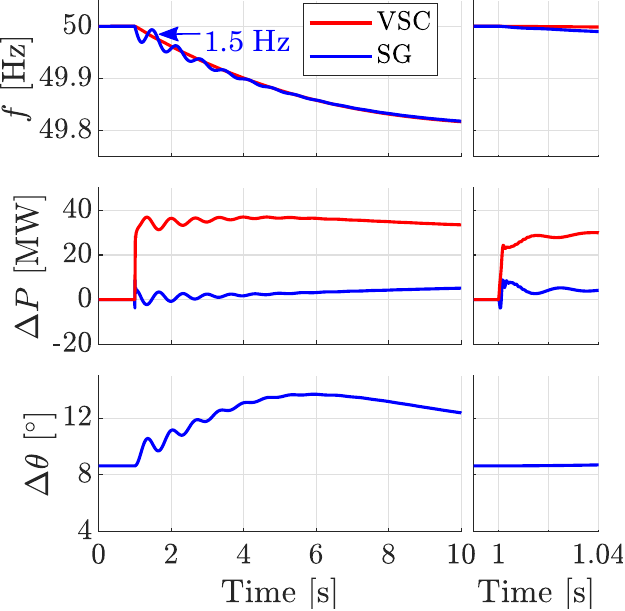}
    \label{fig:sim_results_100_tau_5s}}
    
	\caption{Case 2: Non-linear simulation results.}
	\label{fig:case2_sim_results}
\end{figure*}

Analogously to Case~1, EMT simulations of the non-linear model have been executed to validate the results obtained in the small-signal analysis, whose results are shown in Fig.~\ref{fig:case2_sim_results}. Table~\ref{tab:case_2_freq_chars} gathers the main information about the frequency response. The frequency nadir is similar for any value of $\tau_{p-gfor}$ and higher than those obtained in Case~1, showing that converter's flexibility can help to improve the maximum frequency deviation. It should be noticed that the frequency nadirs in Case~2 are close to the steady-state value after the load change, which is 49.8~Hz, revealing that the frequency dynamics are similar to a first-order response. Regarding the RoCoF, $\tau_{p-gfor}$ can have a big impact on the SG frequency derivative. For low values of $\tau_{p-gfor}$ (10~ms and 100~ms), the RoCoF obtained is higher than in Case~1, 
but are still within the secure operational range of 0.5-1~Hz/s even for such a low-inertia system. For higher values of $\tau_{p-gfor}$, the converter imposes a slower response to the frequency, whose RoCoF is drastically reduced until 0.08~Hz/s for $\tau_{p-gfor}$ = 5~s. The energy required from the converter considerably increases for higher values of $\tau_{p-gfor}$, but it should not be an issue due to its larger size.

Similarly to Case~1, some key findings can also be drawn for this case:
\begin{itemize}
    \item Frequency dynamics in low-inertia systems can be an opportunity to shape its behaviour due to the higher VSC's flexibility. First-order dynamics can be achieved.
    \item This flexibility must be ensured with enough converter capacity and available resource to satisfy the transient needs of the grid.
    \item The system can operate stably even for low values of $\tau_{p-gfor}$. The RoCoF does not exceed the limits defined for actual power systems. 
    \item The frequency deviation can be reduced and controlled, as the nadir matches the steady-state value after the disturbance. This differs from nadir found in second-order responses of conventional SG-based power systems.
    \item Due to the system's inertia reduction and VSC's dominance, the remaining SGs might be more sensitive to suffer from electro-mechanical oscillations during the synchronisation. Therefore, the converter control should be designed to ensure a safe SG's operation.
\end{itemize}

\section{Validation in large power systems}
\begin{figure}[t]
	\centering
	\includegraphics[width=1\linewidth]{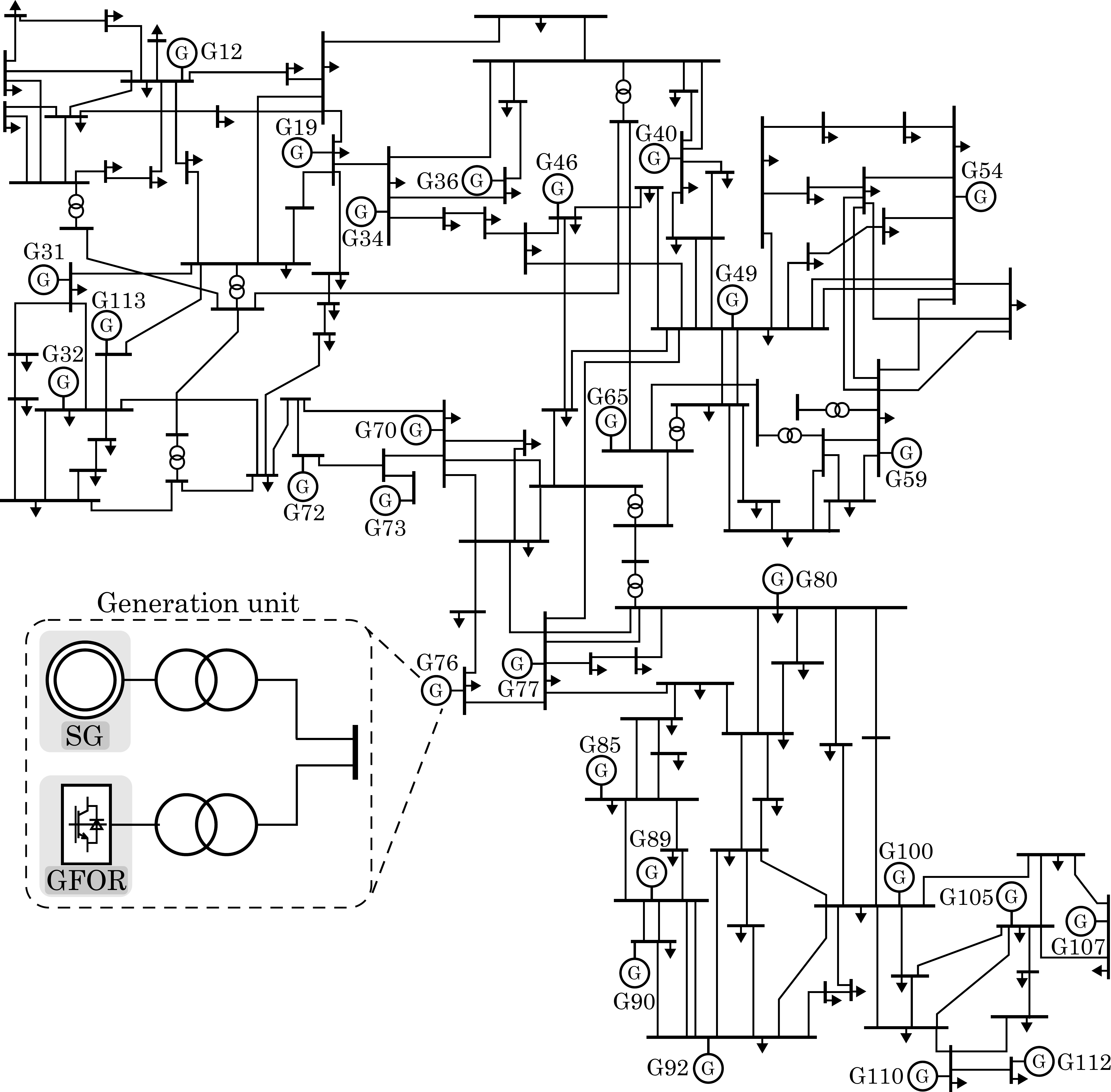}
	\caption{Scheme of the IEEE 118-bus system.}
	\label{fig:system_118_scheme}
\end{figure}

In this section, the previous conclusions obtained in the reduced system are examined for a larger power system with multiple converters and generators. In particular, a modified version of the IEEE 118-bus system has been implemented in PSCAD. Details about the grid parameters and the specific operation point can be found in \cite{PSCAD2018}. This version of the IEEE 118-bus system includes 28 generations units, which consist of one SG and one GFOR VSC connected in parallel through transformers, as shown in Fig.~\ref{fig:system_118_scheme}. The generation units comprise a wide variety of power ratings, from 40~MVA to 610~MVA. The specific SG and VSC capacities are set according to the VSC penetration ($\alpha$), keeping the same rated power for the complete generation unit. The same values of $\alpha$ used in Section~\ref{sec:analysis} have been selected, 20\% and 80\%, while two different values of  $\tau_{p-gfor}$ have been implemented, 100~ms and 1~s. The disturbance applied to the system is the disconnection of the largest generation unit, which is supplying 547~MW, the 14\% of the total generation. 

\begin{figure}[t]
	\centering
	
	\subfloat[$\alpha$ = 20\%; $\tau_{p-gfor}$ = 100 ms]{\includegraphics[width=0.47\linewidth]{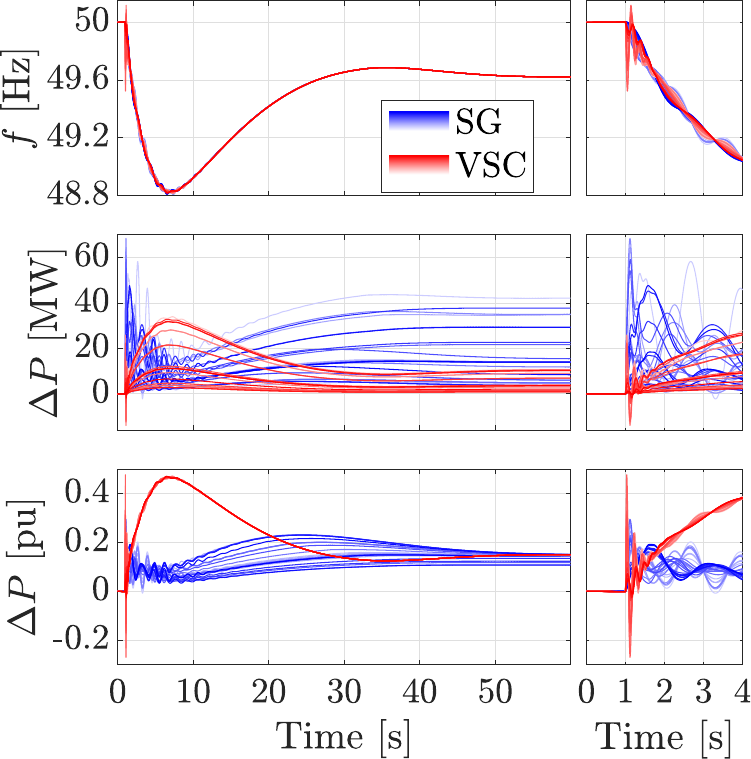}
	\label{fig:sim_resutls_118_20_tau_100ms}}
	\subfloat[$\alpha$ = 20\%; $\tau_{p-gfor}$ = 1 s]{\includegraphics[width=0.47\linewidth]{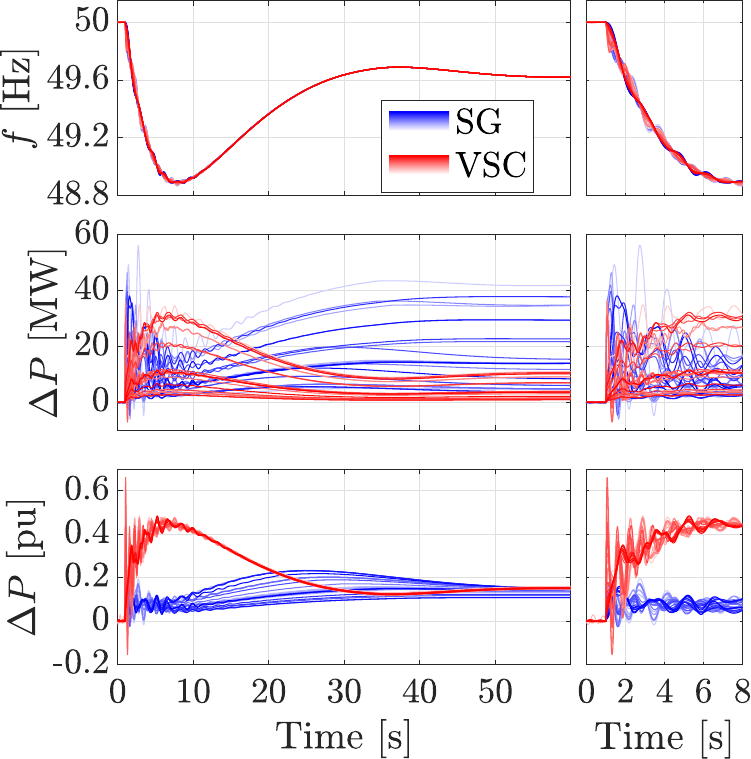}
    \label{fig:sim_resutls_118_20_tau_1s}}
	
	\subfloat[$\alpha$ = 80\%; $\tau_{p-gfor}$ = 100 ms]{\includegraphics[width=0.47\linewidth]{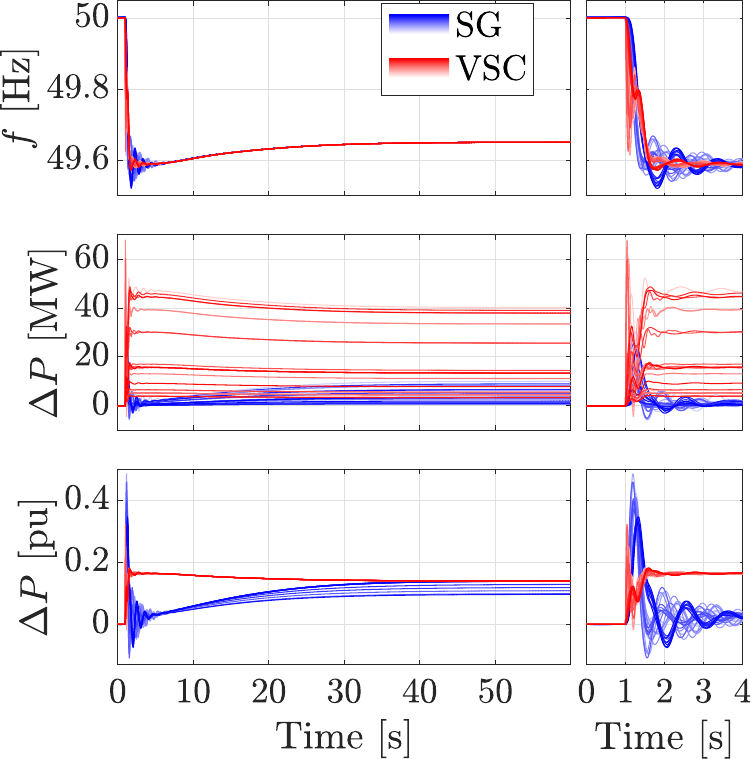}
    \label{fig:sim_results_118_80_tau_100ms}}
    \subfloat[$\alpha$ = 80\%; $\tau_{p-gfor}$ = 1 s]{\includegraphics[width=0.47\linewidth]{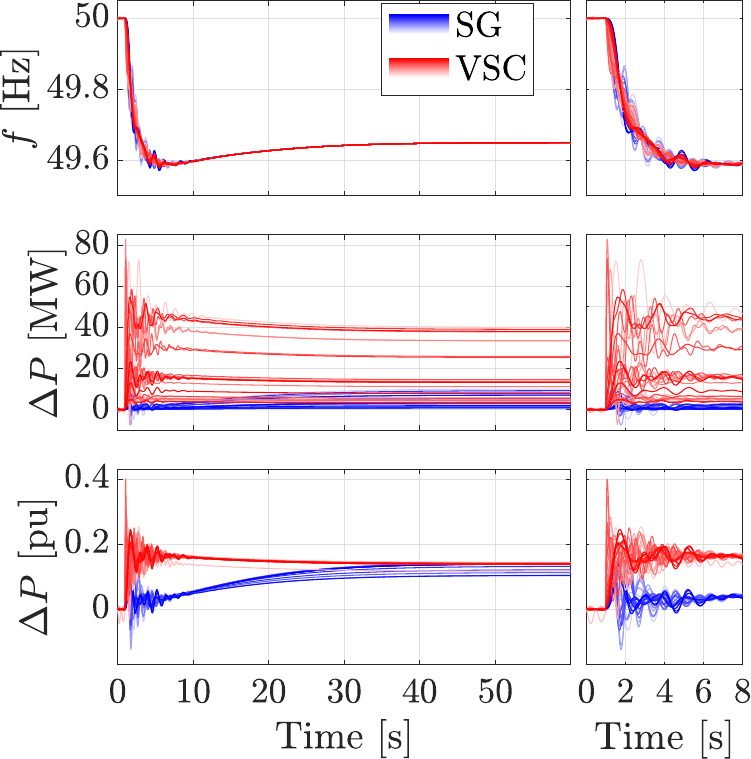}
    \label{fig:sim_results_118_80_tau_1s}}
    
    
	\caption{Non-linear simulation results of the IEEE 118-bus system.}
	\label{fig:sim_results_118}
\end{figure}

Fig.~\ref{fig:sim_results_118} shows the frequency and active power (expressed in MW and pu) of the SGs and VSCs that remain connected. For $\alpha$ = 20\%, second-order frequency responses can be observed for $\tau_{p-gfor}$ equal to 100~ms and 1~s in Figs.~\ref{fig:sim_resutls_118_20_tau_100ms} and~\ref{fig:sim_resutls_118_20_tau_1s}, respectively. As expected, for higher $\tau_{p-gfor}$, the initial RoCoF and frequency nadir are slightly better. However, the synchronisation period is much longer, as observed in the oscillations in active power of the converters. While these oscillations are visible during less than one second for $\tau_{p-gfor}$ = 100~ms, for $\tau_{p-gfor}$ = 1~s they are present in the system for almost 10~s. Despite the different sizes of the converters, their power-frequency characteristics are equivalent, as all they present a 5\% droop. This is observed in the $\Delta P$ expressed in pu, where all the converters show the same response after synchronisation. 

When $\alpha$ is increased to 80\%, the frequency response considerably improves. While the nadir was around 48.8~Hz for $\alpha$ = 20\%, the minimum frequency obtained for $\alpha$ = 80\% is around 49.6~Hz. It can also be observed that the frequency follows first-order dynamics during the first seconds after the disturbance. During this time, the frequency is regulated by the VSCs, as the mechanical torque from the SG turbines is much slower, which finally helps to slightly increase the frequency. Again, a longer synchronisation time is found for $\tau_{p-gfor}$ = 1~s. Therefore, increasing $\tau_{p-gfor}$ might be risky for the system operation, as VSCs become very rigid, forcing the rest units of the system to make the synchronisation task. If all the converters are programmed to respond very slow, the risk of undesired oscillations in the power or synchronisation loss between them might be higher.

\section{Conclusions}

In this paper, the frequency response of power systems with different penetrations of grid-forming converters have been analysed. In particular, a small-signal analysis have been performed in a reduced system consisting of a SG and a GFOR VSC, identifying the main frequency dynamics that could be found for different converter penetrations. Such dynamics can be characterised by two oscillation modes, named here as \textit{Global mode} and \textit{Synchronisation mode}. It has been found that the SG inertia reduction and the higher converter flexibility can help the system to have a more controllable frequency, which can be programmed to have a first-order response. These results have been validated in EMT simulations of a the IEEE 118-bus system, confirming that power systems with high penetration of converters can be an opportunity to shape the frequency dynamics. 

In this scenario with many converters, VSC operation has a great impact on the system performance. Grid-forming operation allows the converter to respond as a voltage source, inherently compensating the generation-load imbalances. As self-synchronised units, the control parameters of the synchronisation loop (droop in this paper) can have an important effect on the converter response. Low values of the time constant in the droop's active power measurement lead to fast synchronisation among converters, reducing the oscillations in the system. Higher time constants can limit the value of the RoCoF, which can be a key aspect for a safe SG operation, but a large amount of energy and power are required from the VSC, which can be beyond the converter's capability. 

Therefore, ensuring enough active power capacity or reserve for grid-forming converters is essential to compensate the load imbalances. It has been already highlighted in the literature that future power systems with high penetrations of renewable will have significant curtailments \cite{Kroposki2017}. However, having such a reserve would be a cost for the renewable plant owner. Specific grid services and market might be studied to stimulate converter-based generation to have such energy and power reserves and therefore providing flexibility through its voltage-source capability to enhance the system frequency dynamics.

\section*{Acknowledgment}
HP2C-DT project. Project TED2021-130351B-C21 is funded by MICIU/AEI/10.13039/501100011033 and by the European Union NextGenerationEU/PRTR.


\ifCLASSOPTIONcaptionsoff
  \newpage
\fi




\bibliographystyle{IEEEtran}
\bibliography{References}

@book{Kundur1994,
   author = {Prabha Kundur},
   publisher = {McGrau-Hill},
   title = {Power System Stability and Control},
   year = {1994},
}

@report{ENTSO-E2017,
   author = {ENTSO-E},
   title = {High Penetration of Power Electronic Interfaced Power Sources ({HPoPEIPS})},
   year = {2017},
}

@report{ENTSO-E2016,
   author = {ENTSO-E},
   title = {Frequency Stability Evaluation Criteria for the Synchronous Zone of Continental Europe},
   year = {2016},
}

@report{ENTSO-E2018,
   author = {ENTSO-E},
   issue = {January},
   title = {Rate of Change of Frequency ({RoCoF}) withstand capability},
}

@article{Collados-Rodriguez2020,
   author = {Carlos Collados-Rodriguez and Marc Cheah-Mane and Eduardo Prieto-Araujo and Oriol Gomis-Bellmunt},
   doi = {10.1109/TPWRD.2019.2959541},
   issn = {19374208},
   issue = {4},
   journal = {IEEE Trans. on Power Delivery},
   pages = {2021-2031},
   publisher = {IEEE},
   title = {Stability analysis of systems with high VSC penetration: where is the limit?},
   volume = {35},
   year = {2020},
}

@report{EirGrid2019,
   author = {EirGrid},
   isbn = {Version 3.5},
   issue = {June},
   title = {EirGrid Grid Code},
   year = {2019},
}

@report{NationalGrid2019,
   author = {{National Grid}},
   title = {{DC0079} Frequency Changes during Large Disturbances and their Impact on the Total System - Report to the Authority},
   url = {www.ofgem.gov.uk},
   year = {2019},
}

@article{Wang2020,
   author = {Xiongfei Wang and Mads Graungaard Taul and Heng Wu and Yicheng Liao and Frede Blaabjerg and Lennart Harnefors},
   doi = {10.1109/ojia.2020.3020392},
   issue = {August},
   journal = {IEEE Open Journal of Industry Applications},
   pages = {115-134},
   title = {Grid-Synchronization Stability of Converter-Based Resources—An Overview},
   volume = {1},
   year = {2020},
}

@web_page{Esios,
   author = {{Red Eléctrica de España}},
   title = {ESIOS. Sistema de información del operador del sistema},
   url = {https://www.esios.ree.es/en},
}

@article{Rosso2021,
   author = {Roberto Rosso and Xiongfei Wang and Marco Liserre and Xiaonan Lu and Soenke Engelken},
   doi = {10.1109/ojia.2021.3074028},
   journal = {IEEE Open Journal of Industry Applications},
   pages = {93-109},
   title = {Grid-Forming Converters: Control Approaches, Grid-Synchronization, and Future Trends—A Review},
   volume = {2},
   year = {2021},
}

@article{Morren2006a,
   author = {Johan Morren and Jan Pierik and Sjoerd W.H. de Haan},
   doi = {10.1016/j.epsr.2005.12.002},
   issn = {03787796},
   issue = {11},
   journal = {Electric Power Systems Research},
   pages = {980-987},
   title = {Inertial response of variable speed wind turbines},
   volume = {76},
   year = {2006},
}

@article{Lasseter2020,
   author = {Robert H. Lasseter and Zhe Chen and Dinesh Pattabiraman},
   doi = {10.1109/JESTPE.2019.2959271},
   issn = {21686785},
   issue = {2},
   journal = {IEEE Journal of Emerging and Selected Topics in Power Electronics},
   pages = {925-935},
   publisher = {IEEE},
   title = {Grid-Forming Inverters: A Critical Asset for the Power Grid},
   volume = {8},
   year = {2020},
}

@article{Tayyebi2020,
   author = {Ali Tayyebi and Dominic Grob and Adolfo Anta and Friederich Kupzog and Florian Dorfler},
   doi = {10.1109/JESTPE.2020.2966524},
   issn = {21686785},
   issue = {2},
   journal = {IEEE Journal of Emerging and Selected Topics in Power Electronics},
   pages = {1004-1018},
   publisher = {IEEE},
   title = {Frequency Stability of Synchronous Machines and Grid-Forming Power Converters},
   volume = {8},
   year = {2020},
}

@article{Hoke2021,
   author = {Andy Hoke and Vahan Gevorgian and Shahil Shah and Przemek Koralewicz and Rick Wallace Kenyon and Benjamin Kroposki},
   doi = {10.1109/MELE.2020.3047169},
   issn = {23255889},
   issue = {1},
   journal = {IEEE Electrification Magazine},
   pages = {74-91},
   title = {Island Power Systems with High Levels of Inverter-Based Resources: Stability and Reliability Challenges},
   volume = {9},
   year = {2021},
}

@article{Sajadi2022,
   author = {Amirhossein Sajadi and Rick Wallace Kenyon and Bri-mathias Hodge},
   doi = {10.1038/s41467-022-30164-3},
   isbn = {4146702230164},
   issue = {2490},
   journal = {Nature Communications},
   pages = {1-12},
   publisher = {Springer US},
   title = {Synchronization in electric power networks with inherent heterogeneity up to 100% inverter-based renewable generation},
   volume = {13},
   year = {2022},
}

@article{Tielens2020,
   author = {Pieter Tielens and Dirk Van Hertem},
   doi = {10.1016/j.rser.2015.11.016},
   issn = {1364-0321},
   issue = {2016},
   journal = {Renewable and Sustainable Energy Reviews},
   pages = {999-1009},
   publisher = {Elsevier},
   title = {The relevance of inertia in power systems},
   volume = {55},
   year = {2020},
}

@article{Kroposki2017,
   author = {Benjamin Kroposki and Brian Johnson and Yingchen Zhang and Vahan Gevorgian and Paul Denholm and Bri Mathias Hodge and Bryan Hannegan},
   doi = {10.1109/MPE.2016.2637122},
   issn = {15407977},
   issue = {2},
   journal = {IEEE Power and Energy Magazine},
   pages = {61-73},
   publisher = {IEEE},
   title = {Achieving a 100\% Renewable Grid: Operating Electric Power Systems with Extremely High Levels of Variable Renewable Energy},
   volume = {15},
   year = {2017},
}

@article{Ashton2015,
   author = {Phillip M Ashton and Christopher S Saunders and Gareth A Taylor and Alex M Carter and Martin E Bradley},
   doi = {10.1109/TPWRS.2014.2333776},
   issn = {08858950},
   issue = {2},
   journal = {IEEE Trans. on Power Syst.},
   pages = {701-709},
   publisher = {IEEE},
   title = {Inertia estimation of the GB power system using synchrophasor measurements},
   volume = {30},
   year = {2015},
}

@report{InternationalRenewableEnergyAgencyIRENA2022,
   author = {IRENA},
   isbn = {9789292603649},
   title = {Renewable Energy and Jobs Annual Review 2022},
   year = {2022},
}

@report{GermanTransmissionSystemOperators2020,
   author = {{German Transmission System Operators}},
   issue = {December},
   title = {Need to develop grid-forming STATCOM systems. Position Paper of the German Transmission System Operators},
   year = {2020},
}

@report{Holttinen2019,
   author = {Hannele Holttinen and Juha Kiviluoma and Thomas Levy and Liu Jun and Peter Borre Eriksen and Nicolaos Cutululis and Vera Silva and Emmanuel Neau and Jan Dobschinski},
   title = {Design and operation of power systems with large amounts of wind power},
   year = {2019},
}

@article{Meng2022,
   author = {Lexuan Meng and Rasool Heydari and Haofeng Bai and Jean-Philippe Hasler and Gunnar Ingestrom and Jan Kheir and Andrew Owens and Jan R Svensson},
   journal = {CIGRE Session Paris 2022},
   title = {Energy Storage Enhanced STATCOM for Secure and Stable Power Grids},
   year = {2022},
}

@article{Page2022,
   author = {Frederick Page and Kazuyori Tahata},
   journal = {CIGRE Session Paris 2022},
   title = {Grid-forming FACTS Systems for Increased Renewable Generation Penetration},
   year = {2022},
}

@report{Denholm2020,
   author = {Paul Denholm and Trieu Mai and Rick Wallace Kenyon and Ben Kroposki and Mark O Malley},
   issue = {May},
   journal = {National Renewable Energy Laboratory (NREL)},
   pages = {48},
   title = {Inertia and the Power Grid : A Guide Without the Spin},
   year = {2020},
}

@web_page{Eirgrid2022,
   author = {Eirgrid},
   title = {Electricity Grid to Run on 75\% Variable Renewable Generation Following Successful Trial},
   url = {https://www.eirgridgroup.com/newsroom/electricity-grid-to-run-o/index.xml},
   year = {2022},
}

@report{GobiernodeEspana.MinisteriodeIndustria2020,
   author = {{Energía y Turismo Gobierno de España. Ministerio de Industria}},
   title = {Plan de desarrollo de la Red de Transporte de Energía Eléctrica.},
   year = {2020},
}

@report{InternationalRenewableEnergyAgencyIRENA2020,
   author = {IRENA},
   isbn = {9789292601690},
   title = {Innovative solutions for 100\% Renewable power in {Sweden}},
   year = {2020},
}

@report{ElectraNet2019,
   author = {ElectraNet},
   issue = {February},
   title = {Addressing the system strength gap in SA},
   year = {2019},
}

@report{PSCAD2018,
   author = {PSCAD},
   title = {{IEEE} 118 Bus System},
   year = {2018},
}

@article{Klein1991,
   author = {M Klein and G J Rogers and P Kundur},
   doi = {10.1109/59.119229},
   issn = {15580679},
   issue = {3},
   journal = {IEEE Trans. on Power Syst.},
   pages = {914-921},
   title = {A fundamental study of inter-area oscillations in power systems},
   volume = {6},
   year = {1991},
}

@article{Gu2018,
   author = {Huajie Gu and Ruifeng Yan and Tapan Kumar Saha},
   issue = {2},
   journal = {IEEE Trans. on Power Syst.},
   pages = {1533-1543},
   publisher = {IEEE},
   title = {Minimum Synchronous Inertia Requirement of Renewable Power Systems},
   volume = {33},
   year = {2018},
}

@article{Ekanayake2004,
   author = {Janaka Ekanayake and Nick Jenkins},
   doi = {10.1109/TEC.2004.827712},
   issn = {08858969},
   issue = {4},
   journal = {IEEE Trans. on Energy Conversion},
   pages = {800-802},
   title = {Comparison of the response of doubly fed and fixed-speed induction generator wind turbines to changes in network frequency},
   volume = {19},
   year = {2004},
}

@report{Migrate2017,
   author = {Taoufik Qoria and Quentin Cossart and Chuanyue Li and Xavier Guillaud and Frederic Colas and François Gruson and Xavier Kestelyn},
   title = {{MIGRATE} project. Deliverable 3.2: Local control and simulation tools for large transmission systems},
   year = {2018},
}

@report{AEMO2021,
   author = {AEMO},
   title = {Frequency and Time Error Monitoring - {Quarter} 3 2021},
   year = {2021},
}

@article{Kenyon2023,
   author = {Rick Wallace Kenyon and Amirhossein Sajadi and Matt Bossart and Andy Hoke and Bri-Mathias Hodge},
   doi = {10.1109/JSYST.2023.3257284},
   issn = {19379234},
   journal = {IEEE Systems Journal},
   month = {4},
   title = {Interactive Power to Frequency Dynamics Between Grid-Forming Inverters and Synchronous Generators in Power Electronics-Dominated Power Systems},
   volume = {Early Access},
   year = {2023},
}

@report{NERC2017,
   author = {NERC},
   city = {Atlanta},
   month = {12},
   pages = {1-47},
   title = {Integrating Inverter-Based Resources into Low Short Circuit Strength Systems},
   year = {2017},
}

@report{CIGRE2016,
   author = {{CIGRE WG B4.62}},
   isbn = {9782858733743},
   publisher = {CIGRÉ},
   title = {Connection of wind farms to weak AC networks},
   year = {2016},
}

@web_page{AEMO2023b,
   author = {AEMO},
   journal = {https://aemo.com.au/en/energy-systems/electricity/national-electricity-market-nem/system-operations/ancillary-services/frequency-and-time-deviation-monitoring},
   title = {Frequency and time deviation monitoring},
   year = {2023},
}
\end{document}